# Antineutrino Science in KamLAND

Atsuto Suzuki

KEK: High Energy Accelerator Research Organization,
Oho 1-1, Tsukuba, 305-0801 Ibaragi, Japan

The primary goal of KamLAND is a search for the oscillation of $\bar{\nu}_e$'s emitted from distant power reactors. The long baseline, typically 180 km, enables KamLAND to address the oscillation solution of the "solar neutrino problem" with $\bar{\nu}_e$'s under laboratory conditions. KamLAND found fewer reactor $\bar{\nu}_e$ events than expected from standard assumptions about $\bar{\nu}_e$ propagation at more than 9σ confidence level (C.L.). The observed energy spectrum disagrees with the expected spectral shape at more than 5σ C.L., and prefers the distortion from neutrino oscillation effects. A three-flavor oscillation analysis of the data from KamLAND and KamLAND + solar neutrino experiments with CPT invariance, yields $\Delta m^2_{21}$ = [$7.54^{+0.19}_{-0.18} \times 10^{-5}$ eV$^2$, $7.53^{+0.19}_{-0.18} \times 10^{-5}$ eV$^2$], $\tan^2\theta_{12}$ = [$0.481^{+0.092}_{-0.080}$, $0.437^{+0.029}_{-0.026}$], and $\sin^2\theta_{13}$ = [$0.010^{+0.033}_{-0.034}$, $0.023^{+0.015}_{-0.015}$]. All solutions to the solar neutrino problem except for the large mixing angle (LMA) region are excluded. KamLAND also demonstrated almost two cycles of the periodic feature expected from neutrino oscillation effects. KamLAND performed the first experimental study of antineutrinos from the Earth's interior so-called geoneutrinos (geo $\bar{\nu}_e$'s), and succeeded in detecting geo $\bar{\nu}_e$'s produced by the decays of $^{238}$U and $^{232}$Th within the Earth. Assuming a chondritic Th/U mass ratio, we obtain $116^{+28}_{-27}$ $\bar{\nu}_e$ events from $^{238}$U and $^{232}$Th, corresponding a geo $\bar{\nu}_e$ flux of $3.4^{+0.8}_{-0.8} \times 10^6$ cm$^{-2}$s$^{-1}$ at the KamLAND location. We evaluate various bulk silicate Earth composition models using the observed geo $\bar{\nu}_e$ rate.

## 1. Introduction

The existence of neutrinos was first postulated in 1930 by W. Pauli as a remedy for the continuous energy spectrum found in experiments on the radioactive *β*-decay of atomic nuclei [1]. A neutrino was introduced to be a weakly interacting particle with a neutral charge, a smaller mass than that of an electron and a spin 1/2. In 1934, E. Fermi developed the theory of *β*-decay process, supposing the new concept of particle creation and annihilation processes [2]. In 1956, the discovery of the neutrino came from detecting the inverse *β*-decay process by F. Reines et al., using one of the Savannah River nuclear reactors [3]. In 1956, T.D. Lee and C.N. Yang proposed the parity-violation in weak interactions [4], although the parity had been assumed to be conserved for a long time. Only one year later in 1957, C. Wu et al. found a definite asymmetry in the angular distribution of electrons emitted in the *β*-decay of polarized Co$^{60}$ nuclei [5]. The parity was proved to be fully violated in *β*-decays. In 1957, Pontecorvo discussed the possibility of neutrino-antineutrino oscillations based on the analogy to an existing phenomenon of the $K^0 \leftrightarrows \bar{K}^0$ oscillations [6]. In 1958, M. Goldhaber et al. measured the neutrino helicity directly and found that the neutrino is left-handed [7]. Such experimental and theoretical progresses revised the Fermi $\beta$-decay theory toward realizing the Lorentz-invariant and universal weak interaction Hamiltonian with V-A forms [8]. Here, neutrinos are assumed as two component massless particles, although there is no evidence for massless neutrinos. B. Pontecorvo [9] and M. Schwartz [10] independently proposed the feasibility of neutrino experiments, using accelerators. The new question came up as to whether the neutrinos emitted in the $\pi \rightarrow \mu$ decay and in the $\beta$-decay are identical or not. In 1962, the experiment at BNL confirmed that the $\pi$–decay neutrino ($\nu_\mu$) is different from the $\beta$-decay neutrino ($\nu_e$) [11]. Based on the existence of two kinds of neutrinos, a particle mixture theory of neutrinos was proposed, which formulates neutrino oscillations of $\nu_e \leftrightarrows \nu_\mu$ with their masses and mixing angles [12]. In 1960's and 1970's, neutrinos were used to probe the structure of nucleons and the property of weak interactions. A key-prediction of the Glashow-Weinberg-Salam model, so-called the Standard Model was the existence of weak neutral current interactions mediated by the $Z^0$ boson in addition to the already known $W^\pm$ charged bosons. In 1973, the Gargamelle collaboration discovered the weak neutral currents in the bubble chamber Gargamelle exposed to the neutrino beam derived from the CERN PS [13]. In 1983, the Super Proton Synchrotron (SPS) at CERN enabled to produce the weak bosons directly. The experimental groups of UA1 led by C. Rubbia and UA2 led by P. Darriulat succeeded in detecting lepton pairs with very large momenta from decays of the $W^\pm$ and $Z^0$ bosons [14]. The precise measurement of the number of light neutrinos ($m_\nu < m_Z/2$) came from studies of $Z^0$ production in $e^+e^-$ collisions. The result was $N_\nu = 2.984 \pm 0.008$ [15]. In 1998, Super-Kamiokande demonstrated the evidence of atmospheric neutrino oscillations [16]. This is the first observation of a finite neutrino mass. The third kind of neutrino called $\nu_\tau$ was observed by DONUT in 2000 [17].

Since 1980's, new large size detectors built in deep underground facilities have substantially contributed to the progress of neutrino physics. These detectors were originally designed for the detection of nucleon decays which were predicted by the idea of Grand Unified Theories. In Japan, the Kamiokande experiment started in 1983, constructing a 3000 ton imaging water Cerenkov detector in a 1000 m underground at the Kamioka mine. In 1987, a neutrino burst from the supernova SN1987A was first detected in Kamiokande [18] and the US experiment IMB [19], which resulted in opening a new research field called the neutrino astronomy. In 1989, Kamiokande succeeded in observing solar neutrinos [20], and confirmed the long-standing puzzle of the solar neutrino deficit which had been first observed by R. Davis et al. almost 20 years ago [21]. In 1992 and 1994, Kamiokande found the atmospheric neutrino anomaly in which the data of the $\nu_\mu/\nu_e$ flux ratio is different from the prediction [22][23]. A gigantic 50,000 ton water Cerenkov detector experiment, Super-Kamiokande started the data-taking in 1996 and solved the atmospheric neutrino anomaly by showing the evidence of neutrino oscillations in 1998 [16]. Neutrino oscillations induced by finite masses and mixing angles became real phenomena and pointed to physics beyond the Standard Model.

The relation between neutrino oscillations and masses is described as follows. Neutrinos participating in the charged current weak interactions are characterized by the flavor ($e, \mu, \tau$). But the neutrinos of a definite flavor are not necessarily states of a definite mass. Instead, they are generally coherent superpositions of such states. For instance, in the two flavor case the states, $|\nu_e\rangle$ and $|\nu_\mu\rangle$ mix with the mass states $|\nu_1\rangle$ and $|\nu_2\rangle$ as $|\nu_e\rangle = \cos\theta_{12}|\nu_1\rangle + \sin\theta_{12}|\nu_2\rangle$, $|\nu_\mu\rangle = -\sin\theta_{12}|\nu_1\rangle + \cos\theta_{12}|\nu_2\rangle$. Neutrino flavor oscillations are a fundamental consequence of two assumptions: that the neutrino has a finite rest mass and that the neutrino flavor eigenstates mix in the mass eigenstates. If a neutrino is initially created in a state of $|\nu_e\rangle$, then the transition probability to $|\nu_\mu\rangle$, at a distance $L$ from the source is

$$P(\nu_e \to \nu_\mu) = \sin^2 2\theta_{12} \sin^2[\Delta m^2_{21} L / 4E_\nu], \qquad (1)$$

where $\Delta m^2_{21} \equiv |m^2_2 - m^2_1|$ is the mass squared difference, and the angle $\theta_{12}$ is known as the vacuum mixing angle.

With the aim of studying neutrino oscillations furthermore, the KamLAND (1000 ton <u>Kam</u>ioka <u>L</u>iquid Scintillator <u>A</u>nti<u>N</u>eutrino <u>D</u>etector) experiment was proposed in 1994 with the aim of detecting the oscillations of electron antineutrino $\bar{\nu}_e$'s emitted from distant power reactors [24]. There are several potential advantages in KamLAND. The Kamioka mine is surrounded with 52 local reactors in 18 Japanese commercial power-stations. Among them, 26 reactors are located at nearly equal distance of 180 kilometer away from the mine and generate a total of ~70 GW ($10^9$ Watt) which corresponds to ~12 % of the world nuclear power-generation. In particular, the Kashiwazaki station shown in Fig. 1 is the world highest power-station with 24.3 GW. The same distance means that the effects of oscillations will add up rather than average out between different reactors. The contribution of the neutrino flux from overseas and Japanese research reactors is less than 5 %. Fig. 1 also shows the map of commercial nuclear power stations in Japan and the expected event rate for one year exposure with a 1000 ton detector as a function of the distance from Kamioka. Applying such desirable conditions as a huge reactor power, an extremely long and definite baseline of 180 km and the relatively lower energy of reactor neutrinos, KamLAND improves the detection sensitivity of the $\Delta m^2$ oscillation parameter by more than 2 orders of magnitude compared to previous reactor experiments. This accessible parameter region covers one of candidate solutions to the solar neutrino deficit problem which is called the Large Mixing Angle (LMA) with $3 \times 10^{-5} < \Delta m^2 < 2 \times 10^{-4}$ (eV$^2$). KamLAND aims at solving the solar neutrino deficit problem under laboratory conditions.

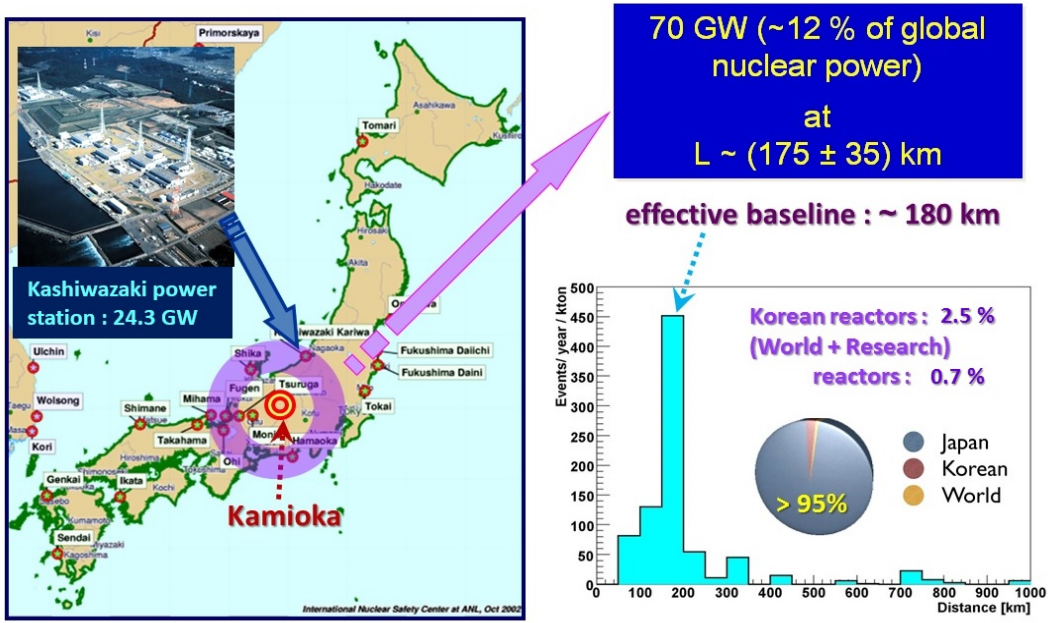

Fig. 1. (Left Panel) Distribution of nuclear power-stations in Japan and some in Korea. (Right panel) Expected neutrino event rate of KamLAND in units of year$^{-1}$ · kton$^{-1}$ from available power plants as a function of the distance from Kamioka.

Additionally, KamLAND is the first detector sensitive to measure the geoneutrinos, $\bar{\nu}_e$'s produced from the $^{238}$U and $^{232}$Th decay chains inside the Earth. One of the basic factors in the interior dynamics and the evolution of the present Earth is the radiogenic heat, ~ 90 % of which comes from the decay of $^{238}$U and $^{232}$Th. Consequently the first detection of geo $\bar{\nu}_e$'s may provide a new window for exploring the deep interior of the Earth.

The success of the experiment depends heavily on how much the background can be suppressed and how many background events can be identified. It is not enough only to make the detector radioactively ultra-pure. To minimize background events, the design must include a high-light-emission liquid scintillator and large aperture photomultiplier tubes (PMT's) with state-of-the-art time and energy response. After proposing the KamLAND project, detector R&D works immediately commenced in particular for developing 17-inch PMT's with high quality performances and a transparent plastic-balloon filled with 1000 tons of liquid scintillator. In 1997, the full budget was funded by the Center Of Excellence (COE) Program sponsored by Japan Society for the Promotion of Science (JSPS). In 1999, 13 US institutes joined the KamLAND project. Since then, the project was performed by the Japan-US collaboration with additional collaborators from China and France afterwards. KamLAND launched into taking data in January 22, 2002.

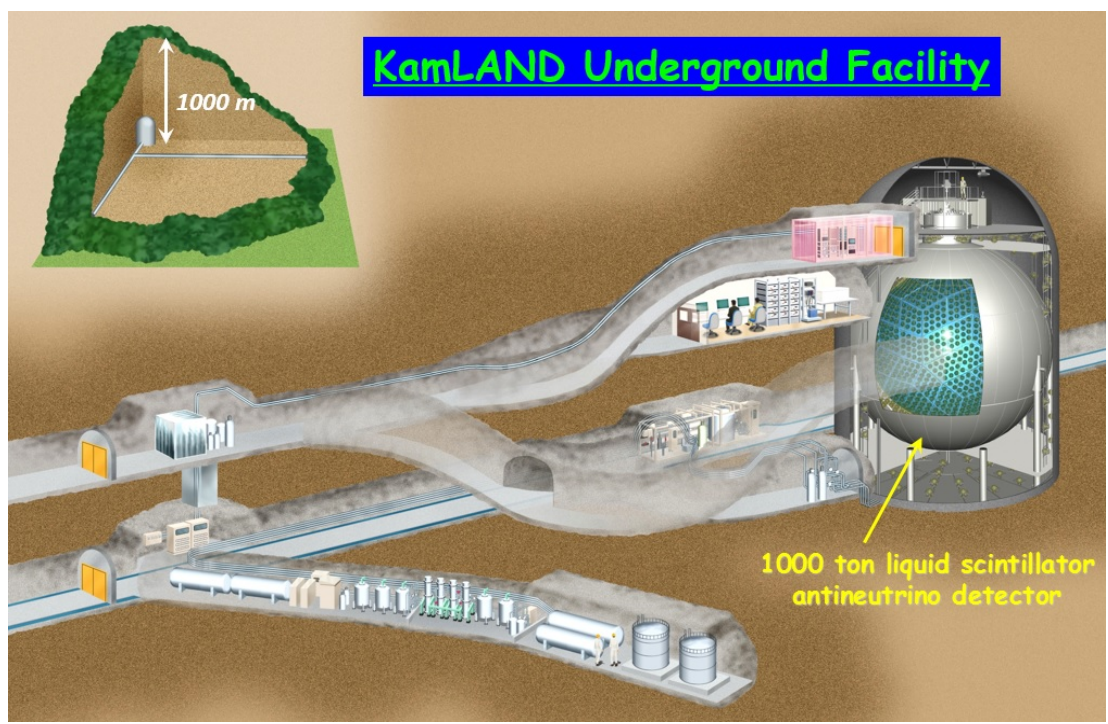

Fig. 2. A bird's-eye view of the KamLAND detector and underground facility.

## 2. KamLAND Detector

KamLAND is built in the Kamioka mine beneath the mountains of Japanese Alps, about 200 km west of Tokyo. The underground laboratory is located 1000 m below the summit of Mt.

Ikenoyama. The detector sits at the site of the old Kamiokande, the 3000 m$^3$ water Cerenkov detector which played a leading role in the study of neutrinos produced via cosmic rays and also helped to pioneer the subject of neutrino astronomy. After dismantling the Kamiokande detector, the rock cavity was enlarged to be 20 m in diameter and 20 m in height. The KamLAND detector consists of a series of concentric spherical shells. Fig. 3 shows a conceptual drawing of the detector.

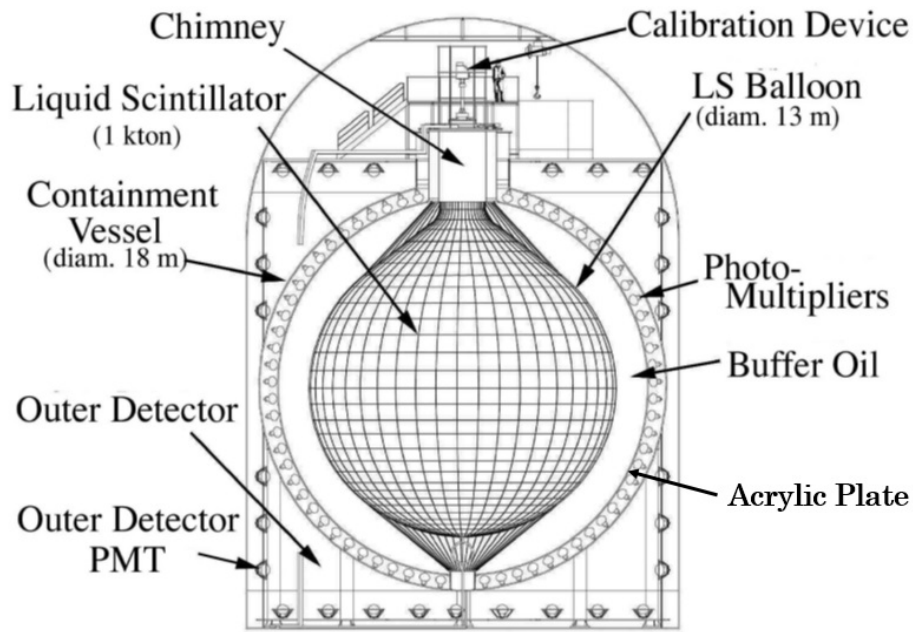

Fig. 3. Schematic view of the KamLAND detector.

The neutrino detector/target is 1000 tons of ultra-pure liquid scintillator located at the center of the detector. The KamLAND liquid scintillator (LS) is a chemical cocktail of 80 % dodecane, 20 % pseudocumene (1,2,4-trimethylbenzene) and 1.36 g/liter of PPO (2,5-diphenyloxazole) as a fluorescence. The light output of the LS is ~8000 photons/MeV. The scintillator is housed in a 13 m-diameter spherical balloon made of 3-layers of nylon with a total thickness of 135 μm and supported by a cargo net structure at the top of the stainless steel vessel. This balloon system hangs inside the 18m-diameter stainless-steel spherical vessel. A buffer mixture of dodecane and isoparaffin oils fills the volume between with the stainless steel vessel and the balloon. Its density is 0.04 % lighter than that of the liquid scintillator to reduce the mechanical load on the balloon. The entire inner surface of the vessel (Inner Detector: ID) is covered by an array of a total of 1879 photomultiplier tubes (PMT's), 1325 of which are specially developed 17-inch and 554 of which are the old Kamiokande 20-inch devices. The total photocathode coverage is 34 %, but only the 17-inch PMT's with 22 %. The 17-inch PMT has the same shape and overall size as those of the 20-inch PMT, but the photosensitive area is restricted to a central 17-inch diameter with an attached black acrylic cover (see Fig. 4(a)). This modification makes it possible to use a box-and-line dynode structure instead of the venetian-blind dynode used in the 20-inch PMT's. As a consequence, under conditions of single-photoelectron illumination at 25 $^o$C and with the applied high-voltage giving a gain of 10$^7$, the 17-inch PMT's offer (1-1.5) ns transit-time spread; output pulse peak-to-valley ratio of (3-5); and a 10 kHz dark count-rate for signals above 1/4 photoelectron. Better than that, the 17-inch tubes show not only a linear response for up to 1000 photoelectron signal level, but also no saturation even at 10,000 photoelectron illumination. This allows us to study more physics associated with events that result from high-energy deposition inside the detector generated by atmospheric neutrinos, nucleon decays, and so on. The comparison of PMT performance between the 17-inch and 20-inch PMT's is shown in Fig. 4 (b).

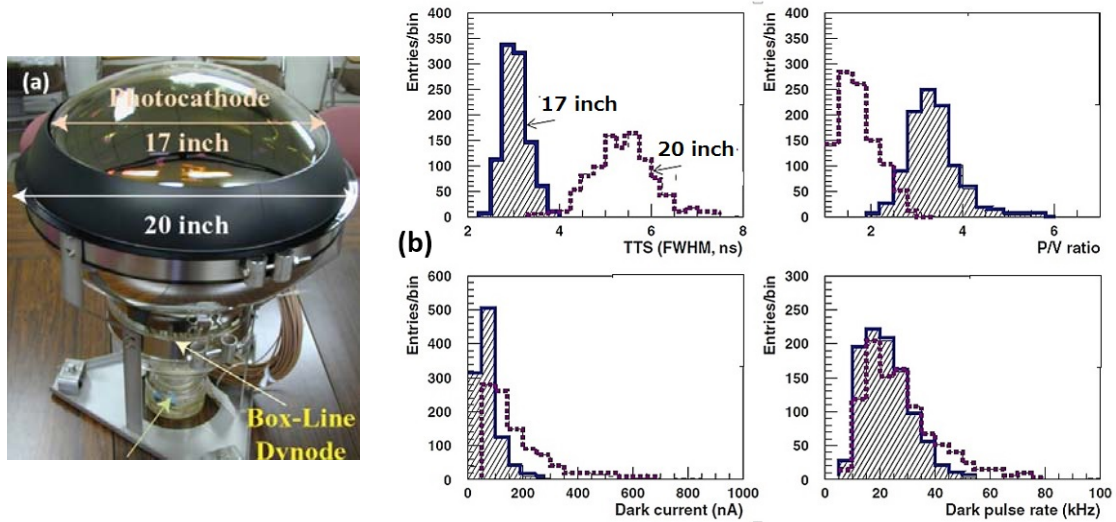

Fig. 4. (a) 17-inch PMT. (b) Performance of the 17-inch PMT (solid histograms), comparing with that of the Super-Kamiokande 20-inch PMT (dotted histograms).

In order to prevent radon emanating from PMT glasses from entering the liquid scintillator, a 3-mm-thick acrylic barrier framed by stainless plates is set in front of the PMT surface (see Fig. 5(c)). The inactive buffer oil serves as passive shielding against external backgrounds such as $\gamma$ rays coming from the PMT glass and nearby rocks. The central detector stands in the cylindrical rock cavity. The volume between the sphere vessel and the cavity is filled with ~3200 m$^3$ of pure water in which 225 Kamiokande 20-inch PMT's are placed to detect cosmic-ray muons by their Cerenkov light. This outer detector (OD) absorbs $\gamma$ rays and neutrons from the surrounding rock and provides a tag for cosmic-ray $\mu$'s. Each PMT signal in ID is recorded, using the analog-transient-waveform-digitizer (ATWD). The ATWD's are self-launching with a threshold ~1/3 photoelectrons and operated with 3 different gains allowing a dynamic range of ~1 mV - 1 V. There are 128 samples per waveform with a sampling time of 1.5 nsec. 2 ATWD sets for each PMT are equipped to reduce detector dead time. The primary ID trigger is set at 200 PMT hits, corresponding to about 0.7 MeV. This threshold is lowered to 120 hits for 1 msec after the primary trigger to detect delayed signals with lower energies. The OD trigger threshold corresponds to > 99 % tagging efficiency. Fig. 5 shows snapshots of the detector construction.

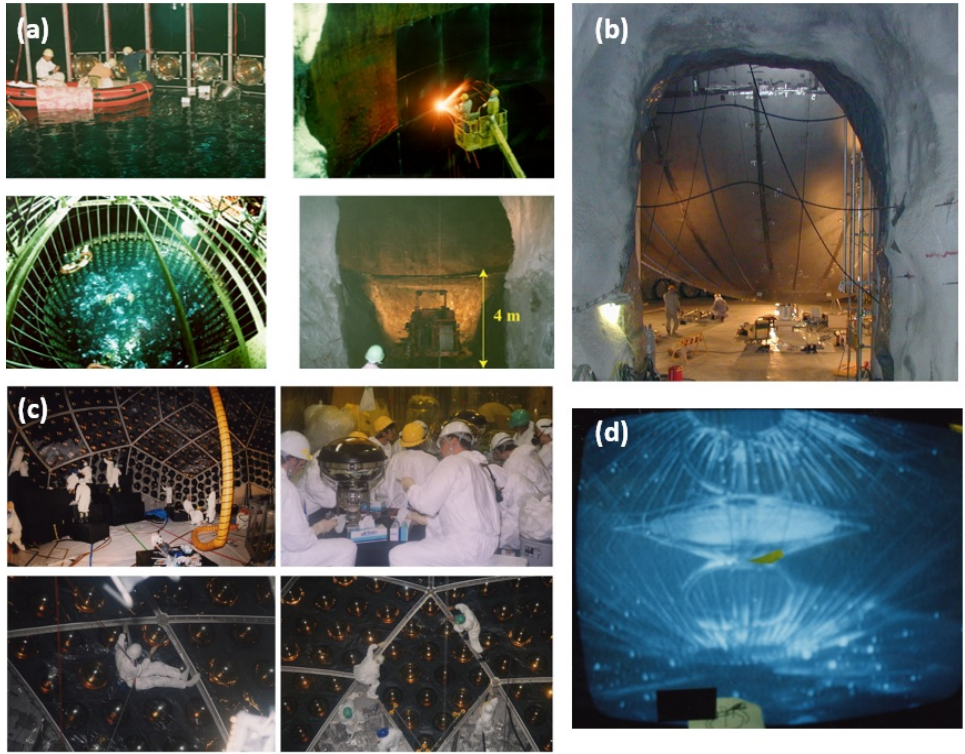

Fig. 5. (a) Kamiokande dismantling in 1998. (b) Stainless steel vessel construction in 1999-2000. (c) PMT installation in 2001 and (d) Oil-fill inside the detector in 2001.

## 3. Detector Performance

The KamLAND detector performance is investigated, using laser and LED light-sources, radioactive sources of $^{203}$Hg ($1\gamma$: 0.279 MeV), $^{68}$Ge ($2\gamma$: 2 × 0.511 MeV), $^{65}$Zn ($1\gamma$: 1.116 MeV), $^{60}$Co ($2\gamma$: 1.173 MeV + 1.333 MeV) and Am-Be ($3\gamma$: 2.20 MeV, 4.40 MeV, 7.60 MeV), the spallation-products $^{12}$B and $^{12}$N produced by energetic cosmic-ray $\mu$'s, and $\gamma$'s generated through cosmic-ray $\mu$-induced thermal neutron captures on $^{1}$H and $^{12}$C. Cosmic-ray µ-induced events provide a monitor to examine the position dependence and time variation of the detector performance, since these events are distributed uniformly in space and time.

The location of interactions inside the detector is determined from PMT hit timing; the energy is obtained from the number of observed photoelectrons after correcting for position and gain variations. Determining the position reconstruction uncertainty is carried out by deploying $\gamma$-ray sources along the vertical axis. Deviations of reconstructed positions from the sources are plotted as a function of the vertical position in Fig. 6. ±5 cm uncertainty is obtained inside the fiducial volume of (-5 m < Z < 5 m). Outside the fiducial volume, the deviation increases due both to a lack of PMT's and to concentration of the balloon-supporting lopes and balloon-welding laps around the top and bottom chimneys. The position resolution for 2.506 MeV $\gamma$ from the $^{60}$Co source is 19 cm. The energy dependence of the position resolution is evaluated to ~30 cm / $\sqrt{E}$ (MeV) for energies up to ~8 MeV.

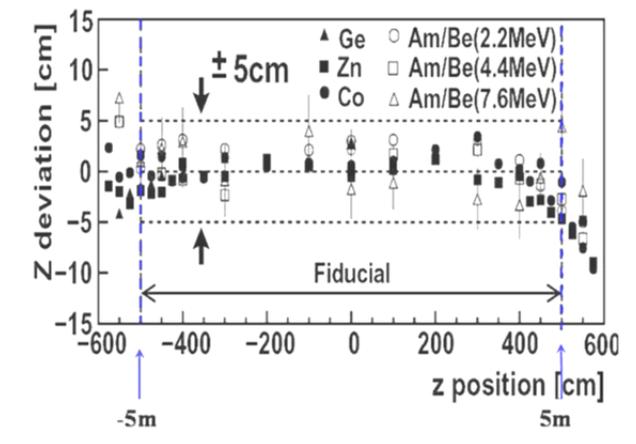

Fig. 6. Deviation of reconstructed vertexes from source positions.

The uncertainties for determining the energy scale come mainly from the non-uniformity in position to position, the time variation due to the detector-operation condition, the non-linearity of the 20-inch PMT response, the additional light yield of Cerenkov light, and the non-linearity of scintillation photons so-called quenching effect. Combining all these effects, the systematic uncertainty in the energy scale at the 2.6 MeV analysis threshold is 2.0 %. The energy resolution is 6.2 % / $\sqrt{E}$ (MeV). Fig. 7 shows the energy spectra of calibration sources and $\Delta E/E$ in the energy range between 0.3 and 10 MeV.

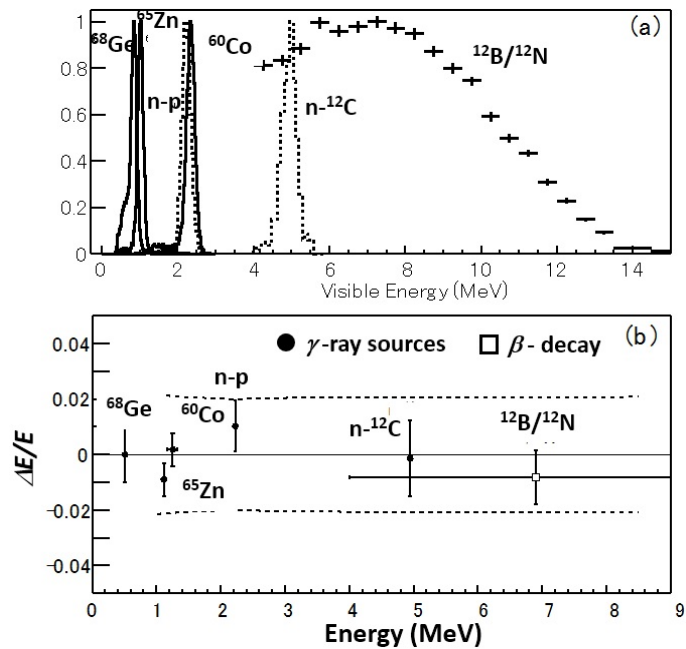

Fig. 7. (a) Energy spectra used in calibration. (b) The fractional difference of the reconstructed average energies and known energies of the source $\gamma$-rays and the $\beta$-rays from $^{12}$B/$^{12}$N.

Radioactive materials inside the liquid scintillator are serious background sources for reactor antineutrino events. The liquid scintillator was purified by the water extraction and gas purging techniques. A Monte Carlo study for reactor neutrino experiments requires that the concentrations of $^{238}$U, $^{232}$Th and $^{40}$K in the liquid scintillator should be lowered to $10^{-13}$ g/g, $10^{-13}$ g/g and $10^{-14}$ g/g. Detecting the sequential chain-decays of $^{214}$Bi → $^{214}$Po → $^{210}$Pb and $^{212}$Bi → $^{212}$Po → $^{208}$Pb are used to estimate the $^{238}$U and $^{232}$Th concentrations. The results are $(3.5 \pm 0.5) \times 10^{-18}$ g/g for $^{238}$U and $(5.2 \pm 0.8) \times 10^{-17}$ g/g for $^{232}$Th. Fig. 8 shows the energy spectra of identified $\beta$'s and $\gamma$'s events in $^{214}$Bi decays into $^{214}$Po and of $\alpha$'s from $^{214}$Po decays.

The $^{40}$K concentration is extracted from the visible energy distribution of single events, subtracting by the contributions from $^{238}$U, $^{232}$Th and $\mu$-induced products. It gives the upper limit of $2.7 \times 10^{-16}$ g/g. These results tell us that the contaminations of $^{238}$U, $^{232}$Th and $^{40}$K inside the liquid scintillator are considerably below the requirements.

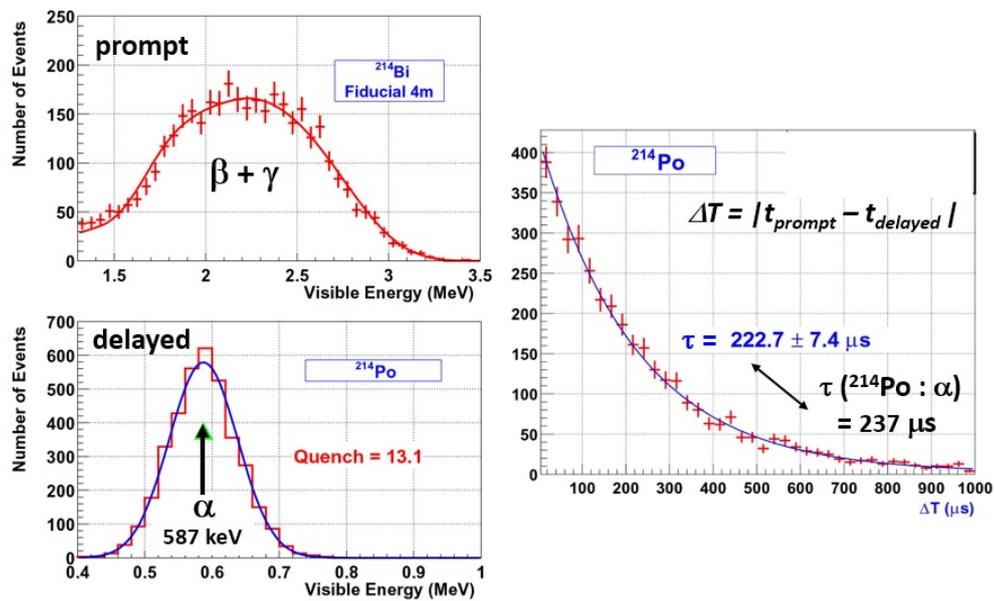

Fig. 8. (Left panel) Visible energy distributions of the prompt $\beta$- and $\gamma$- rays and the delayed $\alpha$-rays in the sequential decay of $^{214}$Bi → $^{214}$Po → $^{210}$Po. (Right panel) The decay time distribution in $^{214}$Po → $^{210}$Po.

An "off-axis" calibration system capable of positioning radioactive sources away from the central vertical axis of the detector was commissioned in 2007 [25]. This calibration system consists of a segmented calibration pole, a variety of radioactive sources, two control cables for the manipulation of the pole inside a glove-box on top of the detector. Fig. 9 (a) illustrates the "off-axis" calibration system. A radioactive source is attached to one end of a pole. It is positioned throughout the fiducial volume by adjusting the orientation and length of the pole. Additional $^{60}$Co pin sources, used for monitoring the pole position, are located along the pole. Fig. 9 (c) is an example of the reconstructed position of the radioactive sources.

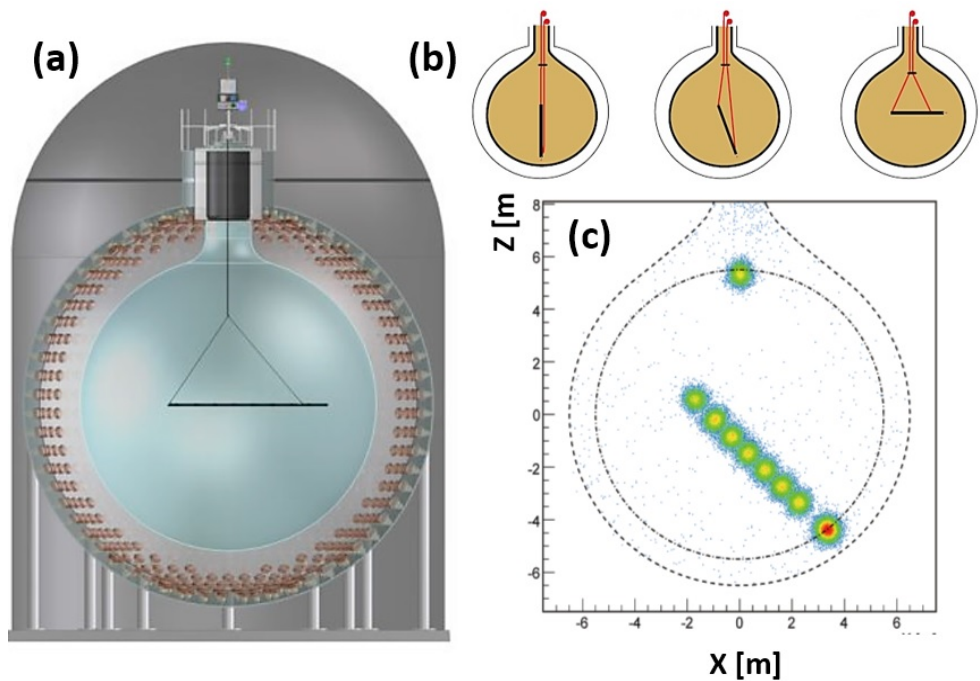

Fig. 9. (a) Illustration of "off-axis" detector calibration system. (b) Manipulation process of control cables. (c) Reconstructed radioactive source position in an azimuthal plane of the detector.

The fiducial volume uncertainty was determined using the measured 3 cm upper limit to the radial deviations. This value yielded a fractional uncertainty in the volume of 1.6 %. A cross-check of this measurement, using cosmic $\mu$-induced spallation events, gave consistent values but with a large uncertainty of 4 %. Thanks to the off-axis calibration system, significant improvement is given in determining the fiducial volume uncertainty. The fiducial volume was extended from 5 m to 5.5 m in radius for the second data sample of reactor neutrino analysis (ANA-II shown in Chapter 6) and 5.5 m to 6.0 m for the third (ANA-III) and fourth (ANA-IV) data sample.

The radial position and energy deviations were measured by varying the source-end with radius and zenith/azimuth angle (see Fig. 10). The measured deviation was found to vary with radius and zenith angle. The magnitudes of the observed systematic deviations are small, < 2 % energy deviation and < 3 cm radial position deviation. They show no significant variation with energy. The variation of these deviations in azimuthal angle is smaller than the variation in radius and zenith angle, as expected from the detector geometry. The off-axis deviations are within the range of earlier estimates, which were deduced from on-axis data and cosmogenic-induced backgrounds.

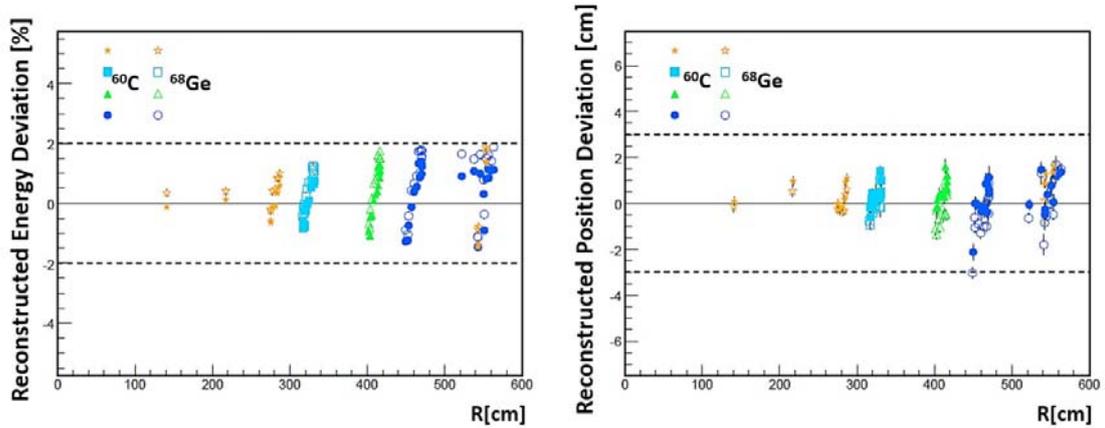

Fig. 10. The measured reconstruction deviations as a function of detector radius: the energy deviations < 2 % and the radial position deviations < 3 %. The different points correspond to a given pole configuration.

## 4. Reactor Neutrinos

Nuclear reactors are very intense sources of an anti-electron-neutrino ($\bar{\nu}_e$) produced through $\beta$-decays of neutron rich fission fragments. Neutrinos from nuclear reactors are more than 99.999 % pure $\bar{\nu}_e$ at $E_\nu$ > 1.8 MeV. Only 4 fissile nuclei of $^{235}$U, $^{238}$U, $^{239}$Pu and $^{241}$Pu dominate the neutrino production and similar energy release from those fissile nuclei ($^{235}$U: 201.7 MeV, $^{238}$U: 205.0 MeV, $^{239}$Pu: 210.0 MeV, $^{241}$Pu: 212.4 MeV) makes a strong correlation between thermal power output and neutrino flux. The neutrino intensity can be roughly estimated to be ~ 2 × $10^{20}$ $\bar{\nu}_e$/GW$_{th}$/sec. Here GW$_{th}$ stands for thermal power output in units of giga-watt.

The information of instantaneous thermal power, fuel burn-up, fuel-exchange and fuel-enrichment records for all Japanese power reactors is required to determine the reactor $\bar{\nu}_e$ flux and to calculate the fission rate for each fissile element. The thermal power generation is checked with the independent records of electric power generation. Figs. 11(a) and (b) show one example of thermal power data and the corresponding fission-rate calculations for fissile elements of $^{235,\,238}$U and $^{239,241}$Pu of which elements contribute to 99.9 % of the $\bar{\nu}_e$ flux generation. The time-integrated fission flux at Kamioka given by these fuel elements in units of fission number/cm$^2$ is plotted also in Fig. 11(c) as a function of the distance between Kamioka and power stations. Here the accumulation time of this data corresponds to the data-taking interval of March 9, 2002 to January 11, 2004. More than 79 % of the total fission flux arises from 26 reactors within the distance of 138-214 km from Kamioka. The flux weighted average distance is equal to 180 km. The relatively narrow band of distances allows KamLAND to be sensitive to the $\bar{\nu}_e$ spectral distortion for certain oscillation parameters. The contribution to the $\bar{\nu}_e$ flux from Korean reactors is estimated to be (2.46 ± 0.25) % based on the reported electric power generation rates. That from other reactors around the world is (0.70 ± 0.35) % on average.

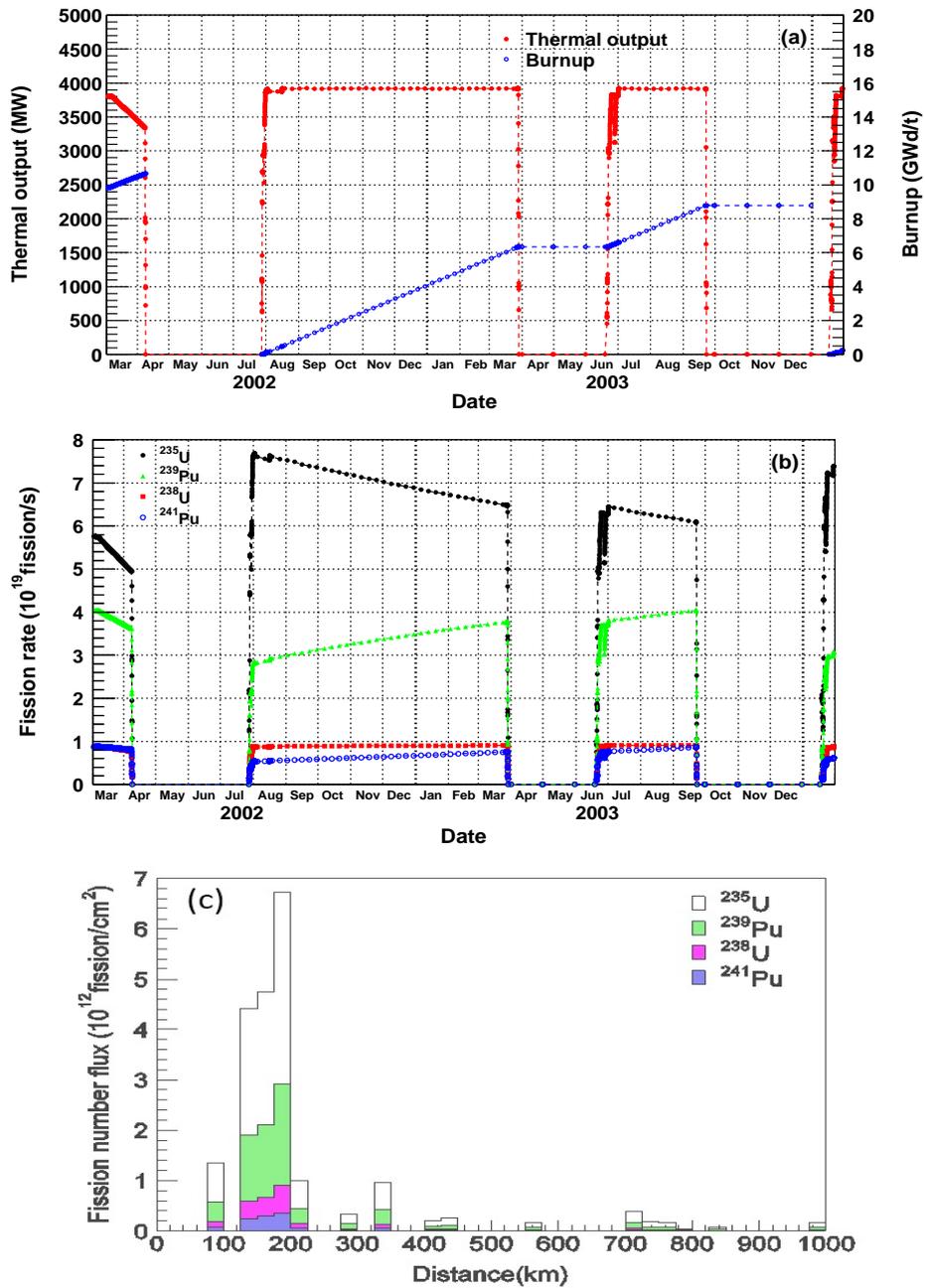

Fig. 11. Example of instantaneous thermal power (a) and fuel burn-up (b) records for one of Japanese commercial reactors. (c) is the fission yields at Kamioka from 4 fissile nuclei. The accumulation period is the same as the data-taking interval of March 9, 2002 to January 11, 2004. These data are provided according to the special agreement between Tohoku Univ. and the Japanese nuclear power-reactor organization.

Although the reactor $\bar{\nu}_e$ flux is calculable in principle, it is very labor-intensive in KamLAND. This is because KamLAND measures reactor neutrinos coming mainly from 53-55 reactors in Japan, and requires detail calculations of the burn-up effect for all reactors. To overcome this difficulty, we developed a simple reactor model [26] so as to accurately calculate the $\bar{\nu}_e$ spectrum of each reactor using the routinely recorded reactor operation parameters. The parameters include the time-dependent thermal output, burn-up and $^{235}$U enrichment of exchanged fuel and its volume ratio.

During the measurement period of KamLAND from March 9, 2002 to January 11, 2004, 52 commercial reactors in 16 electric power stations and a prototype reactor were in operation in Japan. All Japanese commercial reactors are light water reactors (LWR); 29 for boiling water reactors (BWR) and 23 for pressurized water reactors (PWR). Both types of LWR contain 3-5 % enriched Uranium fuel. Generally reactor operation stops once a year for refueling and regular maintenance. During the refueling, one fourth of the total nuclear fuel is exchanged in BWRs and one third in PWR's. To calculate production rates of reactor $\bar{\nu}_e$, knowledge of the correlation between the "core thermal output" and the fission rates is required. The "core thermal output" is defined as the thermal energy generated in the reactor cores, and it is calculated by measuring the heat balance of the reactor cores. The heat taken out by the cooling water is the dominant dissipation source of the reactor energy. Other contributions are less than 1 %. Therefore, the uncertainty of the calculated core thermal output is dominated by the accuracy of measuring the cooling water which itself is given mainly by the accuracy of measuring the flow of the coolant. The accuracy of the flow of the coolant in turn is determined by the uncertainty of the feedwater-flowmeters, which is calibrated to within 2 %. In KamLAND, a value of 2 % is used as the uncertainty of the core thermal output. To calculate the total $\bar{\nu}_e$ flux in KamLAND, it is required to trace the time variation of the fission rate of all reactors, and to understand the burn-up process of nuclear fuel. The process of burn-up is complicated and depends on the core type, history of the burn-up, initial enrichment, fuel exchange history, etc. Detailed simulations exist that calculate the change of the fuel components in accordance with the burn-up. Our simplified model uses only a few reactor operation parameters in calculating $\bar{\nu}_e$ flux, and agrees with the energy spectrum from detailed reactor core simulations within 1 % for different reactor types and burn-up [26]. This simplified model may be applicable to future long-baseline reactor neutrino experiments which made use of several reactors.

Once the fission rates of fissile isotopes are obtained, the $\bar{\nu}_e$ energy spectra except for $^{238}$U are obtained through the following procedure, (i) measurement of the total $\beta$-ray spectrum [27], (ii) fitting with 30 individual hypothetical $\beta$-ray spectra and (iii) conversion of $\beta$-ray spectra to neutrino spectra. Since $^{238}$U undergoes a fast neutron fission, its fission spectrum relies on calculations, considering 744 traces of fission products. Fig. 12(a) shows the neutrino energy spectrum of each fissile isotope in the KamLAND detector. Contributions from long-lived fissile nuclei like $^{106}$Ru ($T_{1/2}$ = 372 days), $^{144}$Ce (285 days) and $^{90}$Sr (28.6 years) in reactor cores and in the cooling pool are not negligible in the low energy region. Although fission spectra reach equilibrium within a day above ~ 2 MeV, the neutrino flux from those nuclei does not have strong correlation with reactor power output and long term average power should be used, instead, to estimate their contributions. Fig. 12(c) shows the expected reactor $\bar{\nu}_e$ energy spectra for four main fissile isotopes at Kamioka.

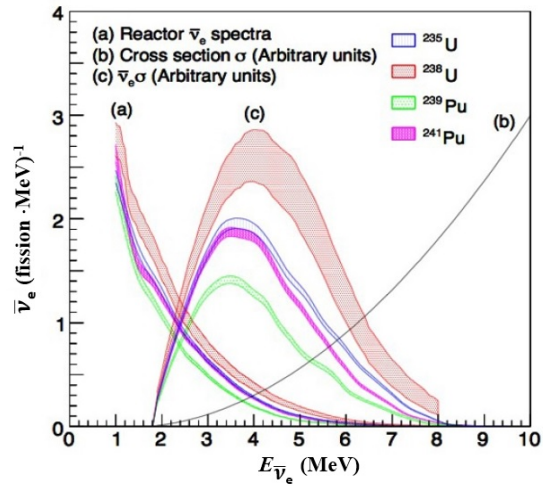

Fig. 12. (a) Reactor $\bar{\nu}_e$ energy spectra for four main fissile isotopes. The shaded region for the isotopes gives the uncertainty in the spectrum. (b) Cross section of the inverse $\beta$-decay reaction. (c) $\bar{\nu}_e$ observed no-oscillation spectrum for each fissile isotope; (c) is a convolution of (a) and (b).

## 5. Reactor Neutrino Detection

Upon entering the detector, $\bar{\nu}_e$ is captured by a free proton and an inverse $\beta$-decay reaction occurs, $\bar{\nu}_e + p \rightarrow e^+ + n$. The positron deposits its energy and then annihilates, yielding two $\gamma$-rays (each 511 keV). The neutron is thermalized in (211.2 ± 2.6) μsec. and then captured by a proton in the following reaction, $n + p \rightarrow d + \gamma$ (2.22 MeV). Thus the inverse β-decay reaction provides a clear sequential signature of the prompt $e^+$ and delayed $\gamma$ with the definite time- and close space-correlations. Although the need to prevent any signals mimicking neutrino events is imperative, these signal correlations give a high rejection-power for background events. Fig. 13 illustrates the inverse $\beta$-decay reaction.

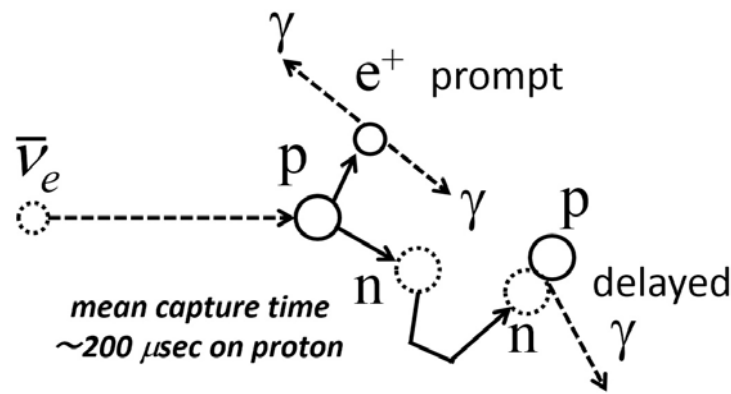

Fig. 13. Sketch of the inverse $\beta$-decay reaction.

The energy threshold of inverse β-decay,

$$E_\nu^{thr} = [(M_n + m_e)^2 - M_p^2] / 2M_p = 1.806 \text{ MeV}, \quad (2)$$

is low enough to observe reactor anti-neutrinos and sufficiently high to avoid major contributions from uncertain long-lived fission nuclei. The neutrino energy $E_\nu$ and the observable positron energy $E_e$ are related by the formula,

$$E_\nu \simeq (E_e + \Delta) \, [1 + E_e / M_e] + (\Delta^2 - m_e^2) / M_p, \quad (3)$$

where $\Delta = M_n - M_p = 1.293$ MeV and the recoil angle is chosen at 90 degree for approximating the average of angular distribution. Roughly $E_\nu \sim E_e + 1.3$ MeV, neglecting the small neutron recoil.

Free neutron decay is an inverse reaction of the anti-neutrino detection reaction and the cross section of the inverse β-decay is related with the neutron lifetime through the formula,

$$\sigma_{tot}^0 = [(2\pi^2 / m_e^5) / (f_{p.s}^R \cdot \tau_n)] \, E_e^0 \cdot p_e^0 \quad (4)$$

Here the phase space factor $f_{p.s}^R = 1.7152$, and $E_e^0 \equiv E_\nu - 1.3$ MeV. The precise measurement of the neutron lifetime with ultra-cold neutrons, $\tau_n = 885.7 \pm 0.8$ sec, greatly improved the precision of the inverse β-decay cross section. Applying order $1/M$ corrections, its precision at relevant energies for reactor neutrino observation (< 10 MeV) is better than ~ 0.2 %. The inverse β-decay cross section is plotted as a function of the neutrino energy in Fig. 12 (b) and the $\bar{\nu}_e$ visible energy spectrum in Fig. 12 (c), convoluting (a) the flux and the cross section (b).

The overall interaction rate and also neutrino spectra models have been experimentally examined with good accuracies. Thanks to previous thorough experiments, current reactor experiments can predict the expected spectrum at a few % levels without any reference detectors close to the reactor cores. The long baseline experiment, KamLAND, observes neutrinos from many country-wide reactor cores. Its successful observation without near detectors became possible thanks to the knowledge from previous efforts [28] for understanding reactor neutrinos.

Nevertheless we can't escape from the geo $\bar{\nu}_e$'s background. KamLAND has the first chance to search for geo $\bar{\nu}_e$'s originated from U/Th decays inside the Earth. The radiogenic heat by U/Th decays plays a dominant role in the energy generation of the Earth. We evaluated the detection rate of geo $\bar{\nu}_e$'s in KamLAND, using various geophysical and geochemical models [29]. In Fig. 14, a smooth broad histogram is the expected visible energy spectrum of positrons produced by reactor $\bar{\nu}_e$'s, and 2 sharp peaks in the energy below 2.5 MeV are expected by geo $\bar{\nu}_e$'s. In the reactor neutrino oscillation analysis, positrons with energies above 2.6 MeV are used to avoid the geoneutrino pollution.

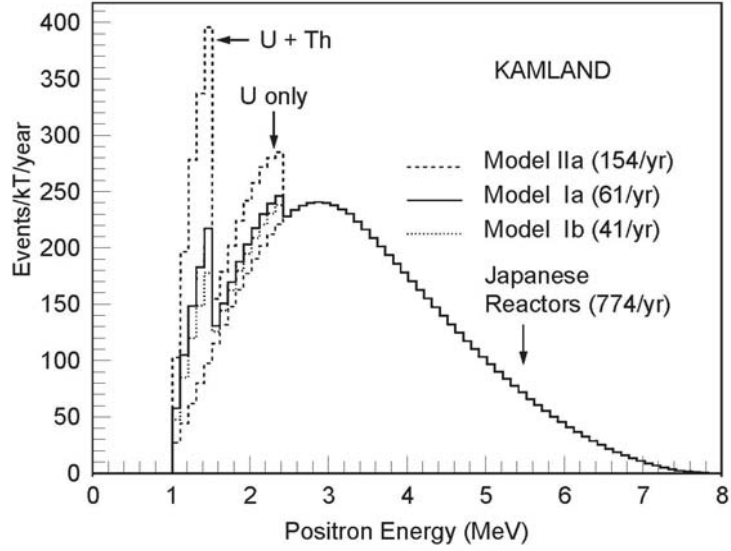

Fig. 14. Expected energy spectra of positrons produced by reactor $\bar{\nu}_e$'s and geo $\bar{\nu}_e$'s.

## 6. Data Analysis

So far KamLAND published 4 papers on the reactor neutrino measurements: "First Results from KamLAND: Evidence for Reactor Anti-neutrino Disappearance", using the first period data sample taken in March 4, 2002 to October 6, 2002 (ANA-I) [30]; "Measurement of Neutrino Oscillation with KamLAND: Evidence of Spectral Distortion" with the second sample taken up to January 11, 2004 (ANA-II) [31]; "Precision Measurement of Neutrino Oscillation Parameters" with the third sample taken up to May 12, 2007 (ANA-III) [32] and "Reactor On-Off Antineutrino Measurement with KamLAND" with the fourth sample taken up to November 20, 2012 (ANA-IV) [33]. These data samples correspond to a total exposure time of 162.2, 766.3, 2881 and 5780 ton-yr., respectively.

**6.1 Event Selection**

$\bar{\nu}_e$'s are detected in KamLAND with the delayed coincidence method for the prompt ($e^+$) and the delayed ($\gamma$) signals in the inverse $\beta$-decay reaction, $\bar{\nu}_e + p \rightarrow e^+ + n$. In ANA-I and -II the analysis uses events with visible energies ($E_{vis}$) more than 2.6 MeV ($E_{vis} \equiv E_{e+} + m_{e+} \sim E_\nu - 0.8$ MeV) in order to avoid the uncertainty of the geoneutrino contribution.

Events with less than 10,000 photoelectrons which corresponds to $\sim$ 30 MeV and no OD (Outer Detector)-tag are categorized as "reactor-$\bar{\nu}_e$ candidates". More energetic events are "cosmic-ray μ candidates". The criteria for the selection of $\bar{\nu}_e$ events in ANA-I are the following: (i) fiducial volume ($R < 5$ m), (ii) time correlation between the prompt $e^+$ and delayed $\gamma$ (0.5 μsec $< \Delta T$ (|prompt - delay|) $< 660$ μsec), (iii) vertex correlation ($\Delta R$ (|prompt - delay|) <1.6 m), (iv) delayed $\gamma$ energy (1.8 $< E_{delayed} < 2.6$ MeV), and (v) a requirement that the delayed vertex position be more than 1.2 m from the central vertical axis to eliminate background from LS monitoring thermometers. The overall efficiency for events from criteria (ii)-(v) including the effect of (i) on the delayed vertex is (78.3 ± 1.6) %. In ANA-II more elaborate selection cuts are used: $R < 5.5$ m, 0.5 μsec $< \Delta T <$ 1000 μsec, $\Delta R < 2$ m and $2.6 < E_{prompt} < 8.5$ MeV. The efficiency of $\bar{\nu}_e$ event selection is improved to (89.8 ± 1.5) %. Fig. 15 shows the vertex distributions of prompt and delayed candidate events observed in ANA-II. Dots in this figure are events without the above selections. These single events dominate around the chimney at the top and the balloon surface. After applying the inverse β-decay event selections, red circles (large dots) remain as the $\bar{\nu}_e$ candidates.

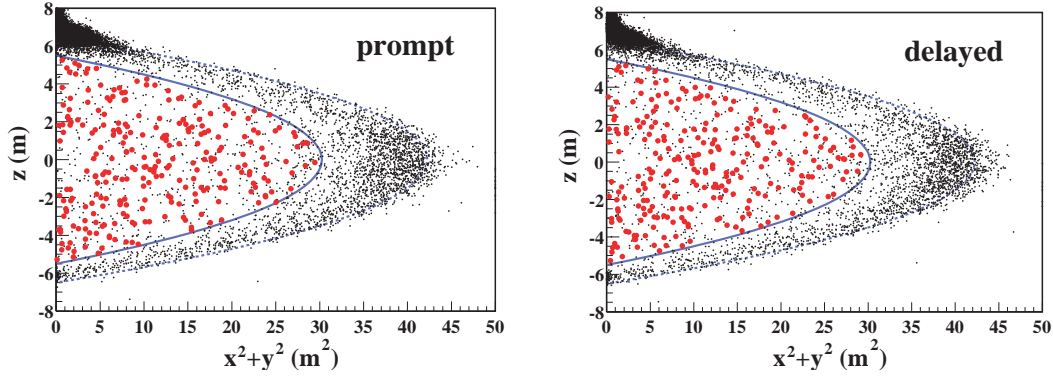

Fig. 15. Vertex distributions of the prompt and delayed events before applying the inverse β-decay event selections. The solid curve stands for the fiducial limit ($R = 5.5$ m) and the dotted curve for the balloon position ($R = 6.5$ m).

The correlation of prompt and delay energies for the ANA-II $\bar{\nu}_e$ candidates before applying the $E_{delayed}$ cut is plotted in Fig. 16. A clear event-isolation in the delayed energy window defined by two dashed lines can be seen. Events concentrated in $E_{delayed} \sim 1$ MeV are expected to be accidental backgrounds. The event rate of $E_{delayed} \sim 5$ MeV is consistent with the expected neutron radiative capture rate on $^{12}C$, and these events are not used in ANA-I and ANA-II due to very low statistics.

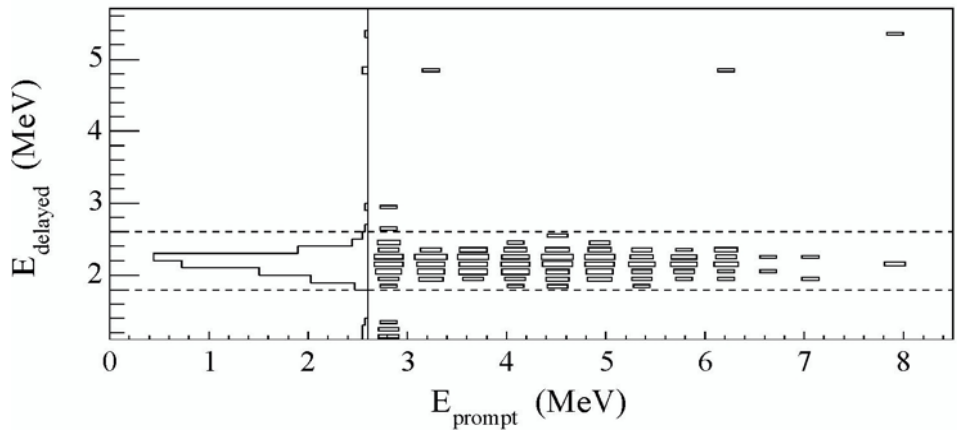

Fig. 16. Scatter plot of $E_{prompt}$ and $E_{delayed}$ for the $\bar{\nu}_e$ candidate events.

The trigger efficiency was determined to be 99.98 % with LED light sources. The combined efficiency of the electronics, data acquisition, and event reconstruction was studied using time distributions of uncorrelated events from calibration $\gamma$ sources. This combined efficiency is better than 99.98 %. The detection efficiency for delayed events from Am-Be source (4.4 MeV prompt $\gamma$ and 2.2 MeV delayed neutron capture $\gamma$ with $\Delta R$ < 1.6 m) was verified to 1 % uncertainty.

The total volume of the liquid scintillator (LS) is 1171 ± 25 m$^3$, as measured by flow meters during filling into the balloon. The nominal 5.5-m-radius fiducial volume ($4\pi R^3/3$) corresponds to 0.595 ± 0.013 of the total LS volume. The effective fiducial volume is defined by cuts on the radial positions of the reconstructed event vertices. In ANA-I and -II, only z-axis calibrations are available, so the systematic uncertainty in the fiducial volume was assessed by studying uniformly distributed cosmic-ray $\mu$ spallation products with the $\beta$-decays of $^{12}$B (Q = 13.4 MeV, $\tau_{1/2}$ = 20.2 msec) and $^{12}$N (Q = 17.3 MeV, $\tau_{1/2}$ = 11.0 msec). The number of $^{12}$B / $^{12}$N events reconstructed in the fiducial volume compared to the total number in the entire LS volume was 0.607 ± 0.006 (stat) ± 0.006 (syst). A consistency check in a similar study of spallation neutrons found the ratio 0.587 ± 0.013 (stat). The $^{12}$B / $^{12}$N events typically have higher energy than reactor $\bar{\nu}_e$ candidates, so an additional systematic error accounts for possible dependence of effective fiducial volume on energy. Comparing the prompt and delayed event positions of delayed-neutron $\beta$-decays of $^9$Li (Q = 13.6 MeV, $\tau_{1/2}$ = 178 msec) and $^8$He (Q = 10.7 MeV, $\tau_{1/2}$ = 119 msec) constrained the variation to 2.7 %. Combining the errors from the LS volume measurements, a 4.7 % systematic error on the fiducial volume was obtained.

Background events passing through the above event-selection criteria and thus embedding inside the inverse $\beta$-decay candidates come dominantly from accidental coincidences, the $^9$Li/$^8$He spallation products and the $\alpha$-decays of the Radon daughter in the LS. The following is the background analysis results for the data sample of ANA-II.

The rate of accidental coincidence increases in the outer region of the fiducial volume, since most background sources are external to the LS. This background is estimated with a 10 msec to 20 sec delayed-coincidence window and by pairing random singles events. This method predicts 2.67 ± 0.02 above 2.6 MeV threshold.

Above 2.6 MeV, neutron and long-lived delayed-neutron emitters are sources of correlated backgrounds. The fast neutrons come from cosmic-ray $\mu$'s missed by the OD or interacting in the rock just outside it. This background is reduced significantly by the OD and several layers of absorbers: the OD itself, the 2.5 m of non-scintillating oil surrounding the LS, and the 1 m of LS outside the fiducial volume. This background contributes fewer than 0.89 events to the data sample.

The ~ 1.5 events/kton/day in the cosmogenic $\beta$-delayed-neutron emitters $^9$Li/$^8$He mimic the $\bar{\nu}_e$ signal. From fits to the decay time and $\beta$-energy spectra, $^9$Li decays are mostly observed. The contribution of $^8$He relative to $^9$Li is less than 15 % at 90 % C.L. For isolated, well tracked $\mu$'s passing through the detector, a 2 sec veto within a 3 m radius cylinder around the track is applied. It is estimated that (4.8 ± 0.9) $^9$Li/$^8$He events remain after cuts.

The most significant background source comes indirectly from the $\alpha$-decays radon daughter of the $^{210}$Po in the liquid scintillator. The signal of the 5.3 MeV $\alpha$ particle is quenched below the threshold, but the secondary reaction $^{13}$C($\alpha$, $n$)$^{16}$O produces events above 2.6 MeV. The natural abundance of $^{13}$C is 1.1 %. Special runs to observe the decay of $^{210}$Po established that there were $(1.47 \pm 0.20) \times 10^9$ $\alpha$–decays during the live time of the ANA-II data taking. The $^{13}$C($\alpha$,n)$^{16}$O reaction results in neutrons with energies up to 7.3 MeV, but most of the scintillation energy spectrum is quenched below 2.6 MeV. In addition, $^{12}$C(n, n')$^{16}$C*, and the 1st and 2nd excited states of $^{16}$O produce signals in coincidence with the scattered neutron but the cross section are not known precisely. Fig. 17 depicts a brief concept of $^{13}$C($\alpha$,n)$^{16}$O reaction. Using the $^{13}$C($\alpha$,n)$^{16}$O reaction cross sections [34], Monte Carlo simulations, and detailed studies of quenching effects to convert the outgoing neutron energy spectrum into a visible energy spectrum, $10.3 \pm 7.1$ events are expected above 2.6 MeV. The $\alpha$-induced background was not considered in the ANA-I analysis and would have contributed $1.9 \pm 1.3$ additional background events.

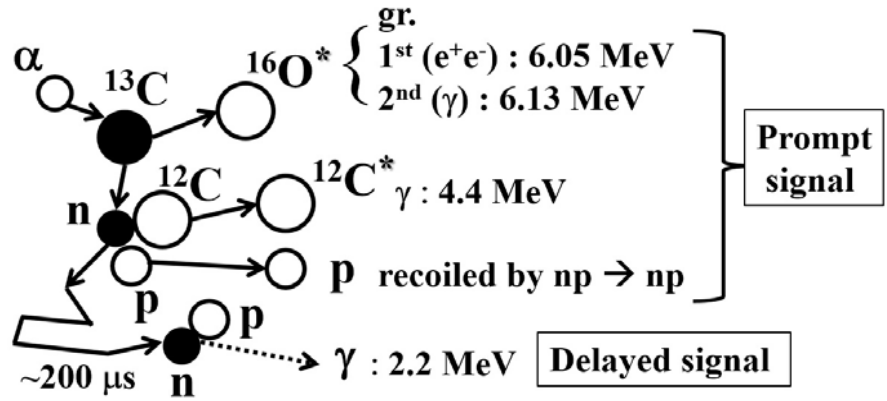

Fig. 17. Illustration of the dominant background process of $^{13}$C($\alpha$, n)$^{16}$O.

Toward precise measurements of reactor neutrinos, the following improvements are essential: (i) reduction of the systematic uncertainties mostly coming from determining the fiducial volume and the absolute energy scale; (ii) understanding the $^{13}$C($\alpha$,n)$^{16}$O reaction well and the quenching phenomena of LS.

A $^{210}$Po$^{13}$C source was developed to study the $^{13}$C($\alpha$,n)$^{16}$O reaction and to tune a simulation using the cross sections from refs. [35] and [36]. The light quenching of the scattered proton by the neutron was precisely measured within $\pm 2$ % by carrying out an experiment using a monochromatic neutron beam to hit the KamLAND LS sample. As a consequence, it is found that the cross sections for the excited $^{16}$O states from ref. [34] agree with the $^{210}$Po$^{13}$C data after scaling the 1st excited state by 0.6; the 2nd excited state requires no scaling. For the ground state, the cross section from ref. [36] and a scale by 1.05 are used in the analysis. Including the $^{210}$Po decay rate assigns an uncertainty of 11 % for the ground state and 20 % for the excited states. Technological efforts to get the above improvements were achieved after the ANA-II analysis and applied to the analysis of the ANA-III and -IV data samples. There should be $10.3 \pm 7.1$, $182.0 \pm 21.7$, $207.1 \pm 26.3$ $^{13}$C($\alpha$,n)$^{16}$O background events in the ANA-II, -III and -IV data samples.

The off-axis calibration system described in the section of "Detector Performance", reduces the fiducial volume uncertainty to 1.6 % inside 5.5 m radius. The position distribution of the $\beta$-decays of $\mu$-induced $^{12}$B/$^{12}$N confirms this with 4.0 % uncertainty by comparing the number of events inside 5.5 m to the number produced in the full LS volume. The $^{12}$B/$^{12}$N event ratio is used to establish the uncertainty between 5.5 m and 6 m, resulting in a combined 6-m-radius fiducial volume uncertainty of 1.8 %.

Off-axis calibration measurements and numerous central-axis deployments of $^{60}$Co, $^{68}$Ge, $^{203}$Hg, $^{65}$Zn, $^{241}$Am$^9$Be, $^{137}$Cs, and $^{210}$Po$^{13}$C radioactive sources established the event reconstruction performance. The vertex reconstruction resolution is ~12 cm / $\sqrt{E}$ (MeV), and the energy resolution is 6.5 % / $\sqrt{E}$ （MeV). The scintillator response is corrected for the nonlinear effects from quenching and Cherenkov light production. The systematic variations of the energy reconstruction over the ANA-III and -IV data samples give absolute energy-scale uncertainties of 1.4 %. Table 1 lists the summary of systematic uncertainties for all data samples. The total systematic uncertainty was reduced to 3.5/4.0 % of ANA-IV from 6.4 % of ANA-I.

| Detector-related | % ANA-I | -II | -III | -IV | Reactor-related | % ANA-I | -II | -III | -IV |
|---|---|---|---|---|---|---|---|---|---|
| Fiducial volume | 4.6 | 4.8 | 2.7 | 1.8/2.5 | $\bar{\nu}_e$ spectra | 2.5 | 2.5 | 2.4 | 1.4 |
| Energy threshold | 2.1 | 2.3 | 1.5 | 1.1/1.3 | Reactor power | 2.0 | 2.1 | 2.1 | 2.1 |
| Efficiency of the cut | 2.1 | 1.6 | 0.6 | 0.7/0.8 | Fuel composition | 1.0 | 1.0 | 1.0 | 1.0 |
| Live time | 0.07 | 0.06 | ----- | ----- | Long-lived nuclei | 0.3 | 0.3 | 0.3 | 0.3/0.4 |
| Cross section | 0.2 | 0.2 | 0.2 | 0.2/0.2 | | | | | |
| Systematic error (%) (ANA-I, -II, -III, -IV) | | | | | | 6.4, | 6.5, | 4.6, | 3.5/4.0 % |

Table 1. Summary of systematic uncertainties relevant to the $\bar{\nu}_e$ event rate. Two values in the ANA-IV show results before/after liquid scintillator purification campaign that continued 2009 [37].

The event selection criteria in ANA-III and ANA-IV were also improved: (i) fiducial volume ($R <$ 6.0 m), (ii) time correlation (0.5 μsec $< \Delta T <$ 1000 μsec), (iii) vertex correlation ($\Delta R <$ 2.0 m), (iv) delayed energy (1.8 $< E_{delayed} <$ 2.6 MeV) or (4.0 $< E_{delayed} <$ 5.8 MeV), corresponding to the neutron-capture $\gamma$ energies for $p$ and $^{12}$C, (v) prompt energy (0.9 MeV $< E_{prompt}\} <$ 8.5 MeV), and (vi) no requirement for eliminating background from LS monitoring thermometers.

## 6.2 Event Rate

Data on the run summary, and the observed, expected and background events are listed, comparing ANA-I, -II, -III and -IV in Table 2.

|  | ANA-I | ANA-II | ANA-III | ANA-IV |
|---|---|---|---|---|
| Exposure (ton-yr) | 162 | 766 | 2881 | 5780 |
| Observed event | 54 | 258 | 1609 | 2611 |
| ($E_{prompt}$ : MeV) | ($E > 2.6$) | ($2.6 < E < 8.5$) | ($0.9 < E < 8.5$) | ($0.9 < E < 8.5$) |
| Expected event | 86.8 ± 5.6 | 365.2 ± 23.7 | 2179 ± 89 | 3564 ± 145 |
| Background event | 0.95 ± 0.99 | 17.5 ± 7.3 | 276.1 ± 23.5 | 364.1 ± 30.5 |
| accidental | 0.0086 ± 0.0005 | 2.69 ±0.02 | 80.5 ±0.1 | 125.5 ±0.1 |
| $^9$Li / $^8$He ($\beta$, $n$) | 0.94 ± 0.85 | 4.8 ± 0.9 | 13.6 ± 1.0 | 31.6 ±1.9 |
| fast neutron | 0 ± 0.5 | < 0.89 | < 9.0 | < 15.3 |
| $^{13}$C ($\alpha$, $n$) $^{16}$O | ------ | 10.3 ± 7.1 | 182.0 ± 21.7 | 207.1 ± 26.3 |

Table 2. Summary of observed and expected events in ANA-I, -II, -III and –IV.

The ratio of observed reactor $\bar{\nu}_e$ events to expected in the absence of neutrino disappearance is 0.611 ± 0.085 (stat) ± 0.041 (syst), 0.658 ± 0.044 (stat) ± 0.047 (syst), 0.593 ± 0.020 (stat) ± 0.026 (syst) and 0.631 ± 0.014 (stat) ± 0.027 (syst) for ANA-I, ANA-II, ANA-III and ANA-IV. KamLAND detected the first evidence for reactor antineutrino disappearance with 99.95 % C.L. in ANA-I and reconfirmed it with 99.998 % C.L. in ANA-II, 8.5 σ C.L. in ANA-III and 10.2 σ C.L. in ANA-IV. Four observations are consistent with each other within the statistical and systematic uncertainties.

The time variation of observed and expected event rates of $\bar{\nu}_e$ candidates is plotted at the 15 data-taking-time-periods of the ANA-IV data sample in Fig. 18. The rates are shown in the two different energy regions of prompt events. One is 0.9 MeV < $E_{prompt}$ < 2.6 MeV, where reactor $\bar{\nu}_e$'s and geo $\bar{\nu}_e$'s coexist. The other is the reactor neutrino dominated region of 2.6 MeV < $E_{prompt}$ < 8.5 MeV. In Fig. 18, the points indicate the measured rates, while the curves show the expected rate variation for reactor $\bar{\nu}_e$'s (black line), reactor $\bar{\nu}_e$'s + backgrounds (blue line) and reactor $\bar{\nu}_e$'s + backgrounds + geo $\bar{\nu}_e$'s (gray line). The contribution of geo $\bar{\nu}_e$'s in 2.6 MeV < $E_{prompt}$ < 8.5 MeV is negligible. Sudden drops of measured event rate in 2007 and 2011 are due to the earthquake which occurred at ~ 70 km north-east away from the Kashiwazaki power station in July 2007 and the major one of March 2011. The measured points agree well with the predictions combined with reactor $\bar{\nu}_e$'s + backgrounds + geo $\bar{\nu}_e$'s in Fig. 18 (a) and reactor $\bar{\nu}_e$'s + backgrounds in Fig. 18 (b). The geo $\bar{\nu}_e$'s rates are calculated from the reference model [38]. The oscillation parameters used to calculate the expected reactor $\bar{\nu}_e$'s rate are the best-fit values from the global oscillation analysis mentioned below (see Table 3).

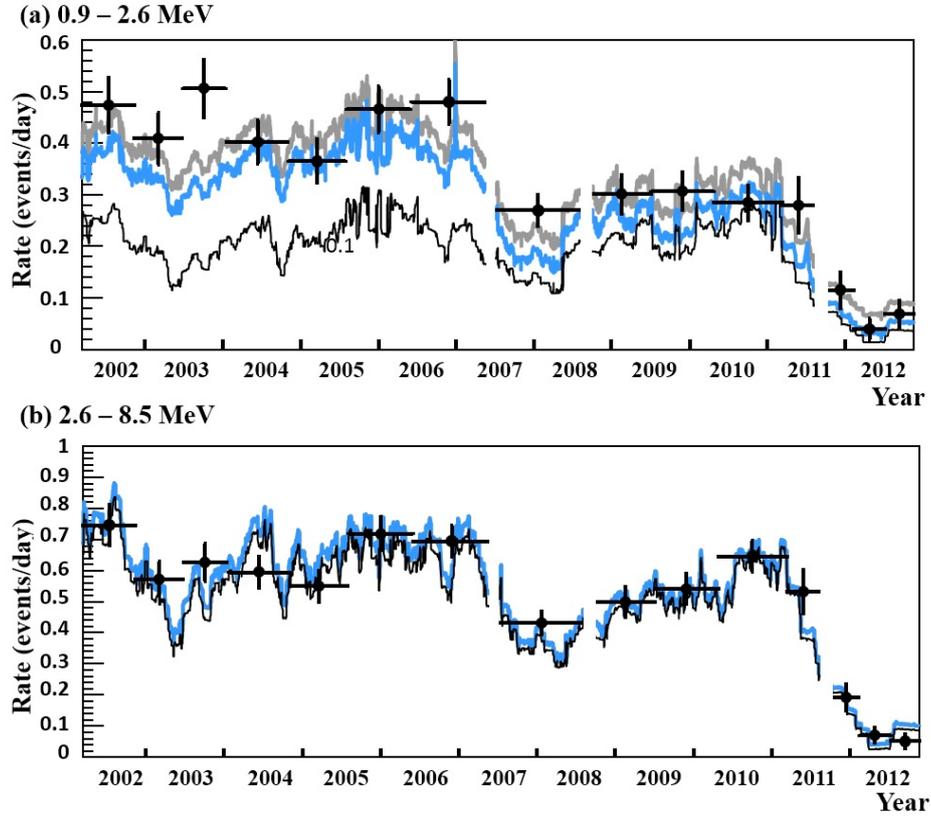

Fig. 18. Time evolution of expected and observed rates at KamLAND for $\bar{\nu}_e$'s with energies between (a) 0.9 MeV and 2.6 MeV and (b) 2.6 MeV and 8.5 MeV. The points indicate the measured rates in a coarse time binning, while the curves show the expected rate variation for reactor $\bar{\nu}_e$'s (black line), reactor $\bar{\nu}_e$'s + backgrounds (blue line) and reactor $\bar{\nu}_e$'s + backgrounds + geo $\bar{\nu}_e$'s (gray line).

Fig. 19 shows the ratio of observed to expected $\bar{\nu}_e$ flux for the ANA-I as well as previous reactor experiments [28] as a function of the average distance from the source. Earlier measurements of reactor neutrinos gave no trace of anomaly. But the first data from KamLAND gives a lower ratio, exactly as expected by one of solar neutrino oscillation solutions (LMA). The dotted curve drawn with $\sin^2 2\theta_{12} = 0.833$ and $\Delta m^2_{21} = 5.5 \times 10^{-5}$ eV$^2$, is representative of a best-fit LMA prediction [39]. Thus the reactor neutrino anomaly prefers the effect expected from neutrino oscillations, assuming CPT invariance.

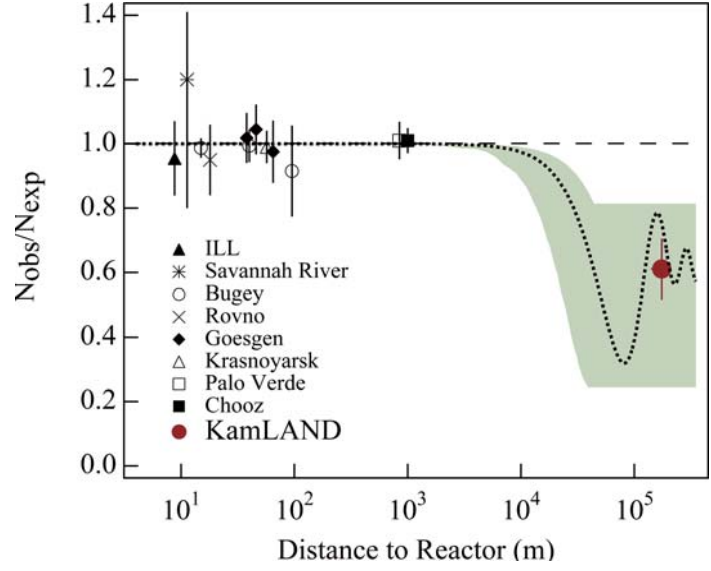

Fig. 19. The ratio of measured to expected $\bar{\nu}_e$ flux from reactor experiments. The solid circle is the KamLAND result plotted as a flux-weighted average distance of ~180 km. The shaded region indicates the range of flux predictions corresponding to the 95 % C.L.

**6.3 Energy Spectrum and Neutrino Oscillation Analysis**

The energy spectrum of reactor neutrinos is uniquely given by convoluting the reactor neutrino flux and the inverse $\beta$-decay cross section. Even if reactor power outputs are less informative, the shape of energy spectrum is reliable. This means that the deformation of reactor neutrino energy spectrum would indicate the existence of additional new physics. Neutrino oscillations characterize the deformation of neutrino energy spectrum depending on oscillation parameters of $\Delta m^2$ and mixing angle $\theta$.

A two-flavor oscillation analysis (with $\theta_{13} = 0$) for the KamLAND event rate and spectrum shape data is carried out, using a maximum likelihood method to obtain the optimum set of oscillation parameters with the following $\chi^2$ definition:

$$\chi^2 = \chi^2_{\text{rate}}(\sin^2 2\theta_{12}, \Delta m^2_{21}, N_{\text{BG1~2}}, \alpha_{1~4})$$
$$-2\ln L_{\text{shape}}(\sin^2 2\theta_{12}, \Delta m^2_{21}, N_{\text{BG1~2}}, \alpha_{1~4})$$
$$+ \chi^2_{\text{BG}}(N_{\text{BG1~2}}) + \chi^2_{\text{distortion}}(\alpha_{1~4}). \qquad (5)$$

$L_{\text{shape}}$ is the likelihood function for the spectrum including experimental distortions. $N_{\text{BG1~2}}$ are the estimated $^9$Li and $^8$He backgrounds and $\alpha_{1~4}$ are parameters to account for the spectral effects of energy scale uncertainty, finite resolution, $\bar{\nu}_e$ spectrum uncertainty, and fiducial volume systematic error, respectively. Parameters are varied to minimize the $\chi^2$ at each pair of $(\Delta m^2_{21}, \sin^2 2\theta_{12})$ with a bound from $\chi^2_{\text{BG}}(N_{\text{BG1~2}})$ and $\chi^2_{\text{distortion}}(\alpha_{1~4})$. With the assumption of CPT invariance, $\bar{\nu}_e$ oscillations are equivalent to $\nu_e$ oscillations. Results from solar neutrino oscillation analysis are also compared in the analysis.

The energy spectra of prompt events ($E_{prompt} \sim E_\nu - 0.8$ MeV) observed in ANA-I and ANA-II are shown in Fig. 20 by dots, together with the expected reactor neutrino spectra without oscillations (black-solid histograms). Associated background spectra are also drawn in the ANA-II figure. In both figures, the data points clearly deviate from the reactor neutrino spectra. The best-fit spectra together with the backgrounds based on two flavor neutrino oscillations (see the next paragraph) are shown by the blue solid histogram in ANA-I and the blue dotted histogram in ANA-II. In the ANA-II analysis, the best-fit of the scaled reactor spectrum (blue-solid histogram) disagrees with the observation, being excluded at the 99.6 % C.L. Here the spectral distortion due to the systematic uncertainties for the backgrounds is considered in the following: for the $^{13}C(\alpha, n)^{16}O$ background shown by the green color in the figure, a free-scale uncertainty around 6 MeV and a 32 % scale uncertainty of the estimated rate around 2.6 MeV and 4.4 MeV are applied to fitting. KamLAND gave the first evidence of the spectrum distortion in neutrino experiments with the confidence level ~ 3 σ. The reactor neutrino anomaly defined as the combined effect of the rate disappearance and spectrum distortion is found at the high confidence level of 99.999995 %.

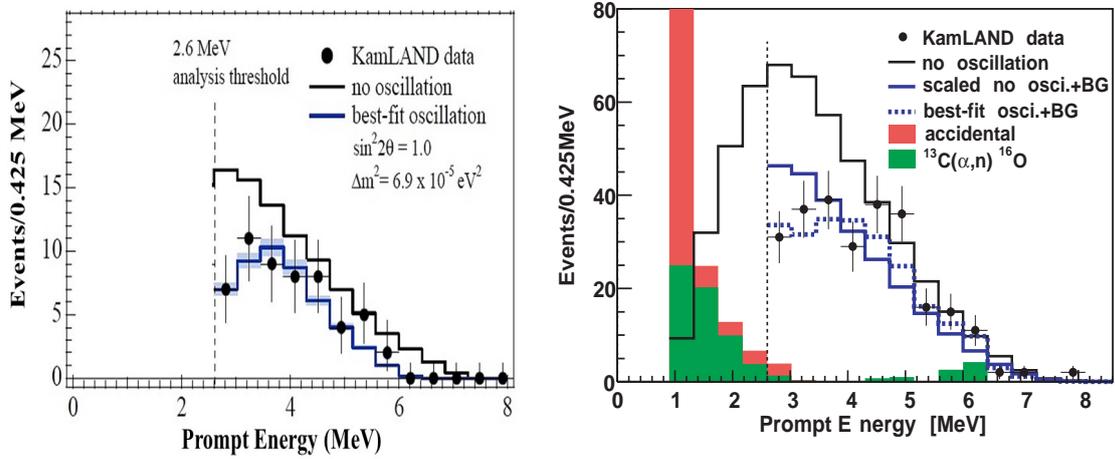

Fig. 20. Energy Spectrum of observed prompt events in ANA-I (Upper panel) and ANA-II (Lower panel). The associated background spectra are also plotted in ANA-II. Histograms are described in the text.

The ANA-I sample is evaluated to be $\sin^2 2\theta_{12} = 1.0$ and $\Delta m^2_{21} = 6.9 \times 10^{-5}$ eV$^2$. These numbers can be compared to the best fit LMA values of $\sin^2 2\theta_{12} = 0.83$ and $\Delta m^2_{21} = 5.5 \times 10^{-5}$ eV$^2$ from [39]. The constraint of oscillation parameters at 95 % C.L. from the ANA-I analysis is shown in Fig. 21(Left panel), combined with the 95 % C.L. allowed region (red color) of the large mixing angle (LMA) solution of solar neutrino experiments [28] and 95 % C.L. excluded regions from CHOOZ (yellow color) [40] and Palo Verde (Green) [41]. In the ANA-II data sample, we account for the $^9$Li accidental and the $^{13}C(\alpha, n)^{16}O$, background rates. The best fit for the rate-shape analysis is $\Delta m^2_{21} = 7.9^{+0.6}_{-0.5} \times 10^{-5}$ eV$^2$ and $\tan^2 \theta_{12} = 0.46$, with a large uncertainty on $\tan^2 \theta_{12}$. A shape-only analysis gives $\Delta m^2_{21} = (8.0 \pm 0.5) \times 10^{-5}$ eV$^2$ and $\tan^2 \theta_{12} = 0.76$. The allowed region contours in $\Delta m^2_{21}$ - $\tan^2 \theta_{12}$ parameter space is shown in Fig. 21 (Right panel), which is derived from the $\Delta \chi^2$ values (e.g., $\Delta \chi^2 < 5.99$ for 95% C.L). The best-fit point is in the region commonly characterized as LMA.

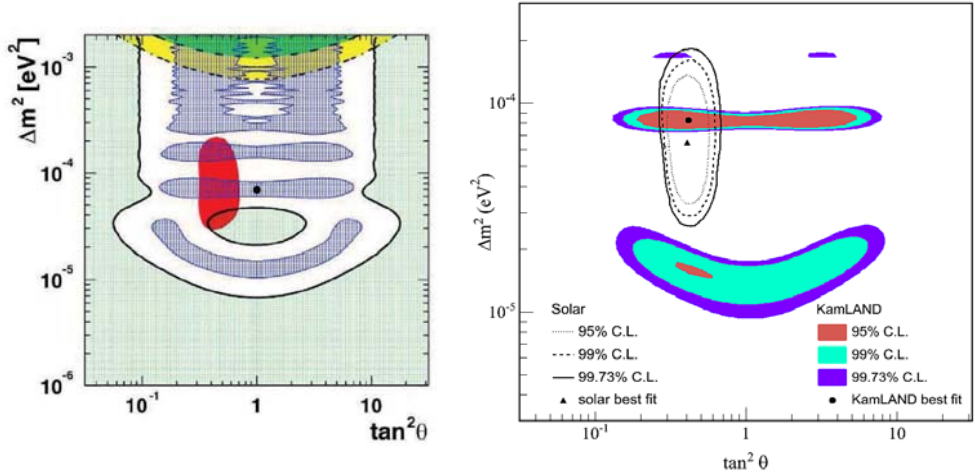

Fig. 21. (Left panel) Excluded regions for the rate analysis and allowed regions for the combined rate and shape analysis from ANA-I at 95 % C.L. The large mixing angle (LMA) solution of solar neutrino experiments is also shown by the red color. (Right panel) Allowed regions at 95 % C.L. from ANA-II (shaded color regions) and solar neutrino experiments (lines) [42].

In the ANA-III analysis, the prompt energy range expands to the geoneutrino energy region. Fig. 22 shows the prompt energy spectrum of selected $\bar{\nu}_e$ events and the fitted backgrounds. The unbinned data are assessed with a maximum likelihood fit to two-flavor neutrino oscillation, simultaneously fitting the geo $\bar{\nu}_e$'s contribution. The method incorporates the absolute time of the event and accounts for time variations in the reactor flux. Earth-matter oscillation effects for geo $\bar{\nu}_e$'s are included. The best-fit to reactor $\bar{\nu}_e$'s oscillations + backgrounds + best-fit geo $\bar{\nu}_e$'s is shown by the blue-solid histogram in Fig. 22 with $\Delta m^2_{21} = 7.58\,^{+0.14}_{-0.13}$ (stat) $^{+0.15}_{-0.15}$ (syst) $\times 10^{-5}$ eV$^2$ and $\tan^2\theta_{12} = 0.56\,^{+0.10}_{-0.07}$ (stat) $^{+0.10}_{-0.06}$ (syst) for $\tan^2\theta_{12} < 1$. Now the scaled reactor spectrum with no distortion from neutrino oscillation is excluded at more than 5 σ.

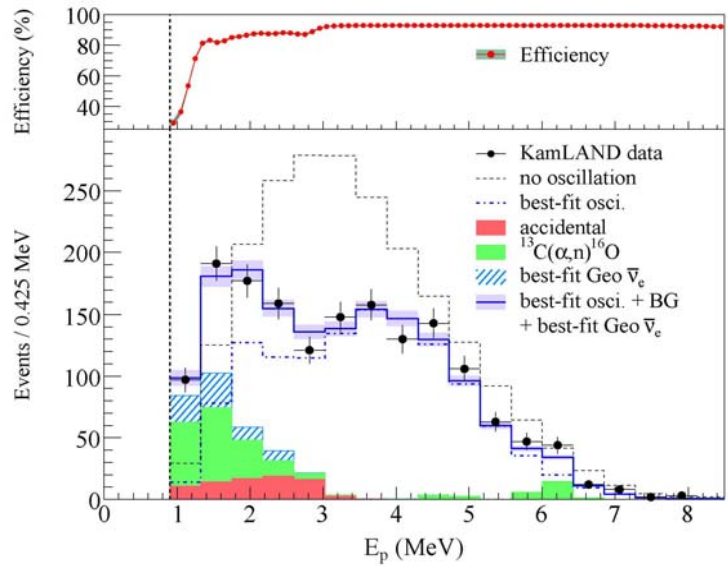

Fig. 22. Prompt event energy spectrum of $\bar{\nu}_e$ candidate events in ANA-III. All histograms corresponding to reactor spectra and expected backgrounds incorporate the energy-dependent selection efficiency (Upper panel). The shaded background and geo $\bar{\nu}_e$'s histograms are cumulative. Statistical uncertainties are shown for the data; the band on the blue histogram indicates the event rate systematic uncertainty.

The allowed contours, including $\Delta\chi^2$ – profiles, are shown in Fig. 23 (Upper panel). Only the best-fit region in the ANA-II analysis remains, while other regions previously allowed at 2.2 σ are disfavored at more than 4 σ. The $\Delta\chi^2$ distributions in this figure tell that the sensitivity in $\Delta m^2$ is dominated by the observed distortion in the KamLAND spectrum, while solar neutrino data provide the best constraint on $\theta$. A global analysis of data from KamLAND, solar neutrino experiments [43] and solar flux experiments [44] gives $\Delta m^2_{21} = 7.59^{+0.21}_{-0.21} \times 10^{-5}$ eV$^2$ and $\tan^2\theta_{12} = 0.47^{+0.06}_{-0.05}$. A stringent constraint to the oscillation parameters is shown in Fig. 23 (Lower panel).

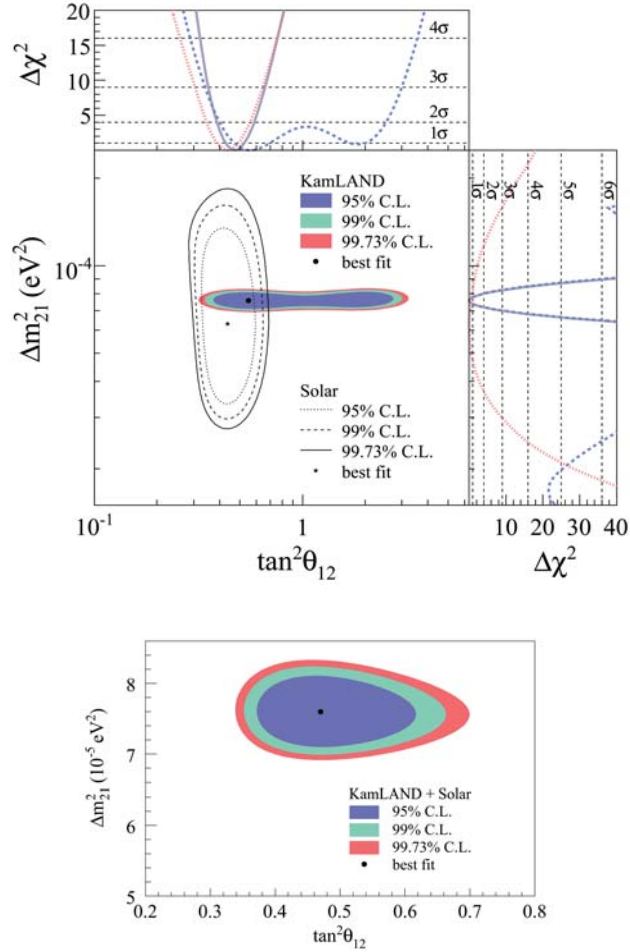

Fig. 23. (Upper panel) Allowed region for neutrino oscillation parameters from ANA-III and solar neutrino experiments. The side-panels show the $\Delta\chi^2$-profiles for KamLAND (dashed line), solar experiments (dotted line) and the combination of the two (solid line). (Lower panel) Allowed region from ANA-III and solar neutrino experiments.

In the ANA-IV sample, three neutrino generations are considered. The $\bar{\nu}_e$ survival probability depends on two mixing angles $\theta_{12}$ and $\theta_{13}$. The $\chi^2$ is defined in the range of 0.9 MeV $< E_{prompt} <$ 8.5 MeV by

$$\chi^2 = \chi^2_{rate}(\theta_{12}, \theta_{13}, \Delta m^2_{21}, N_{BG1\sim5}, N^{geo}_{U,Th}, \alpha_{1\sim4})$$
$$- 2\ln L_{shape}(\theta_{12}, \theta_{13}, \Delta m^2_{21}, N_{BG1\sim5}, N^{geo}_{U,Th}, \alpha_{1\sim4})$$
$$+ \chi^2_{BG}(N_{BG1\sim5}) + \chi^2_{syst}(\alpha_{1\sim4}) + \chi^2_{osci}(\theta_{12}, \theta_{13}, \Delta m^2_{21}). \quad (6)$$

The terms are, in order: the $\chi^2$ contribution for (i) the time-varying event rate, (ii) the time-varying prompt energy spectrum shape, (iii) a penalty term for backgrounds, (iv) a penalty term for systematic uncertainties, and (v) a penalty term for the oscillation parameters. $N^{geo}_{U,Th}$ are the flux normalization parameters for U and Th geo $\bar{\nu}_e$'s, and allow for an Earth-model-independent analysis. $N_{BG1\sim5}$ are the expected number of backgrounds which includes accidental, $^9$Li/$^8$He, $^{13}$C($\alpha$, n)$^{16}$O$_{ground\ state}$, $^{13}$C($\alpha$, n)$^{16}$O$_{excited\ state}$, fast neutron and atmospheric neutrino backgrounds. These terms are allowed to vary in the fit but are constrained with the penalty term (iii). $\alpha_{1\sim4}$ are described before. The penalty term (v) optionally provides a constraint on the neutrino oscillation parameters from solar [45-49], accelerator (T2K [50], MINOS [51]), and short-baseline reactor neutrino experiments (Double CHOOZ [52], Daya Bay [53], RENO [54]). Fig. 24 shows the prompt energy spectra of $\bar{\nu}_e$ candidate events.

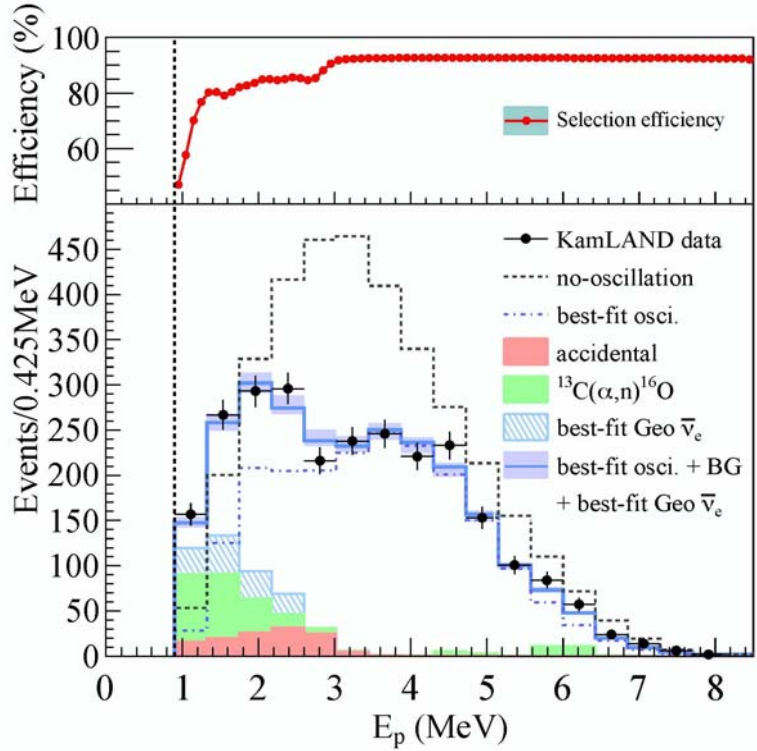

Fig. 24. Prompt energy spectrum of $\bar{\nu}_e$ candidate events above 0.9 MeV energy threshold (vertical dashed line) for the ANA-IV sample.

The fit oscillation parameter values are given for the three-flavor KamLAND-only analysis ($\chi^2_{\text{osci}} = 0$), the combined analysis with solar neutrino experiments and a global analysis also including constraints on $\theta_{13}$ from accelerators and short-baseline reactor experiments, are listed in Table 3.

| Data combination | $\Delta m^2_{21}$ ($10^{-5}$ eV$^2$) | $\tan^2\theta_{12}$ | $\sin^2\theta_{13}$ |
|---|---|---|---|
| KamLAND | $7.54^{+0.19}_{-0.18}$ | $0.481^{+0.092}_{-0.080}$ | $0.010^{+0.033}_{-0.034}$ |
| KamLAND + solar | $7.53^{+0.19}_{-0.18}$ | $0.437^{+0.029}_{-0.026}$ | $0.023^{+0.015}_{-0.015}$ |
| KamLAND + solar + $\theta_{13}$ | $7.53^{+0.18}_{-0.18}$ | $0.436^{+0.029}_{-0.025}$ | $0.023^{+0.002}_{-0.002}$ |

Table 3. Summary of the fit values for $\Delta m^2_{21}$, $\tan^2\theta_{12}$, and $\sin^2\theta_{13}$ from three-flavor neutrino oscillation analyses with various combinations of experimental data.

The extracted confidence intervals in the ($\tan^2\theta_{12}$, $\Delta m^2_{21}$) plane with and without the $\theta_{13}$ constraint are shown in Fig. 25 together with the $\Delta\chi^2$ profiles onto the $\tan^2\theta_{12}$ and $\Delta m^2_{21}$ axes.

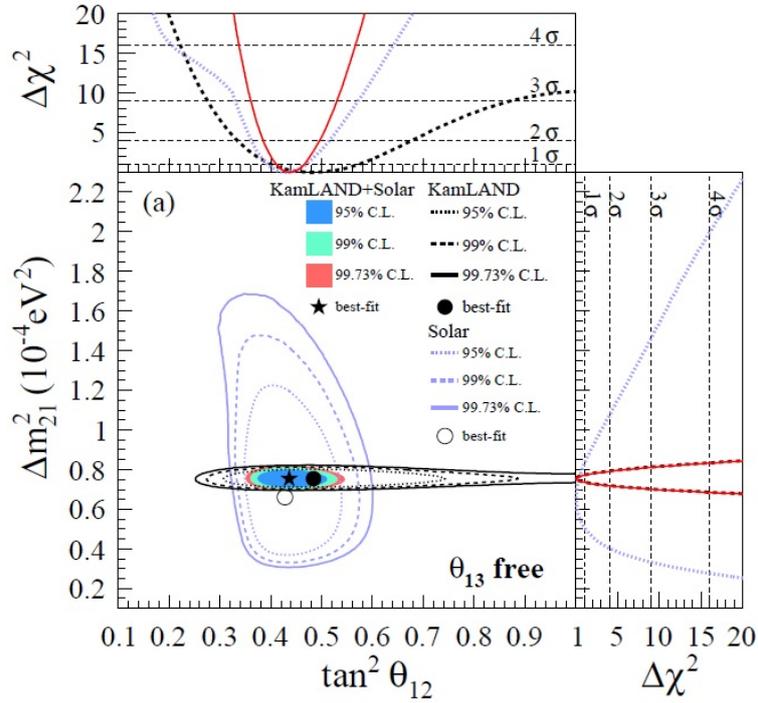

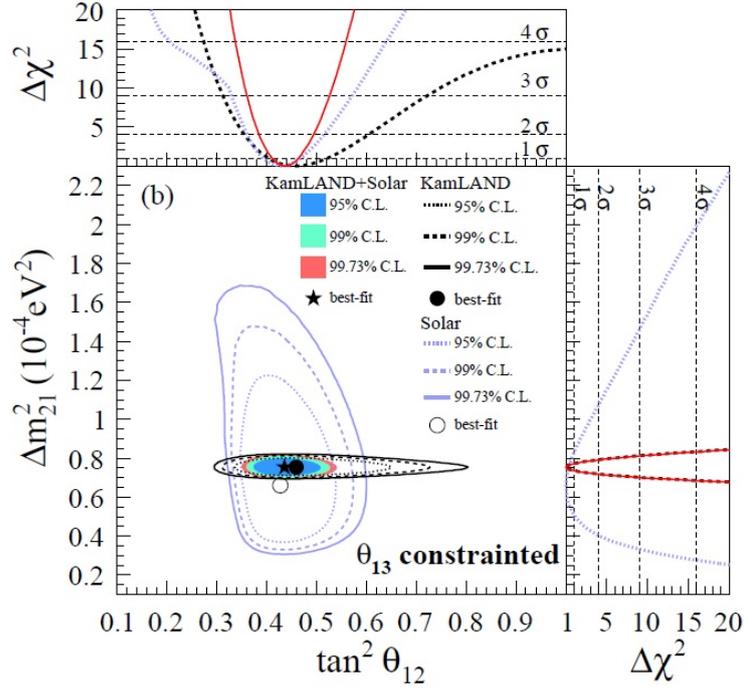

Fig. 25. (Color) Allowed regions projected in the ($\tan^2\theta_{12}$, $\Delta m^2_{21}$) plane, solar and KamLAND data from the three-flavor oscillation analysis for (a) $\theta_{13}$ free and (b) $\theta_{13}$ constrained by accelerator and short-baseline reactor experiments. The shaded regions are from the combined analysis of the solar and KamLAND data.

To illustrate oscillatory behavior of the ANA-III data, the $L_0/E$ distribution is plotted in Fig. 26, where the observed and the best-fit energy spectra are divided by the expected no-oscillation spectrum, including the subtraction of backgrounds and geoneutrinos. $L_0$ is the effective baseline taken as a flux weighted average ($L_0 = 180$ km). The points from the previous reactor experiments are also shown here [28]. The dashed curve in this figure is derived, using the best-fit oscillation parameters with the assumption that all Japanese nuclear reactors are 180 km distant from Kamioka. The histogram and curve show the expectation accounting for the distances to the individual reactors, time-dependent flux variations, and efficiencies. The spectrum exhibits almost two cycles of the periodic feature expected from neutrino oscillations. Two alternative hypotheses for neutrino disappearance: neutrino decay [55] and decoherence [56], give different $L_0/E$ dependences. According to the goodness-of-fit procedure, the decay has a goodness of fit of only 0.7 % ($\chi^2$/d.o.f. = 46.39/16), while decoherence has a goodness of fit of 1.8 % ($\chi^2$/d.o.f. = 52.94/16). Thus, the decay and decoherence hypothesis are excluded at the 3.9 σ C.L. and 4.5 σ C.L. The dash and dash-dot curves in Fig. 27 show the best-fit decay and decoherence behaviors. KamLAND demonstrates the oscillatory shape of reactor $\bar{\nu}_e$'s arising from the neutrino oscillation.

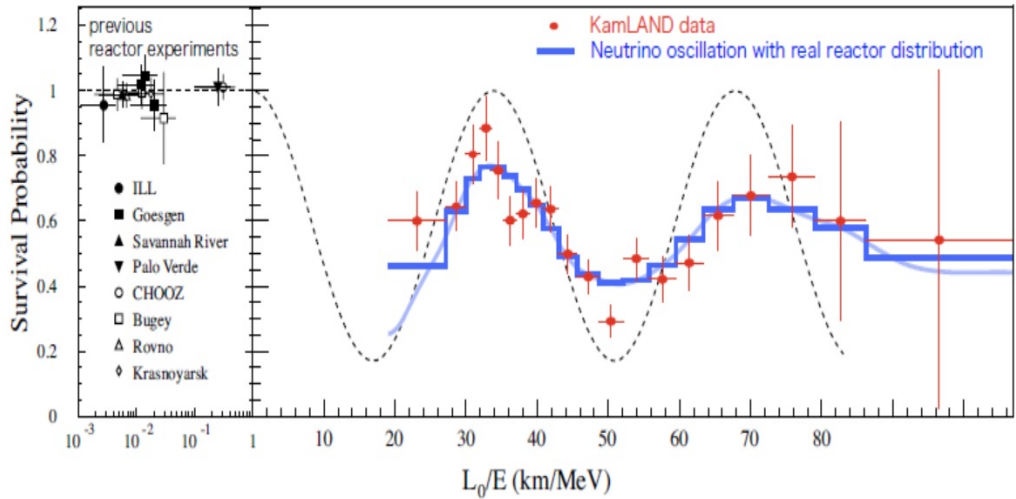

Fig. 26. Ratio of the background and geoneutrino subtracted $\bar{\nu}_e$ spectrum to the expectation for the no-oscillation as a function of $L_0$ (= 180 km$\}$)/$E$. The energy bins are equal probability bins of the best fit including all background (see Fig. 22).

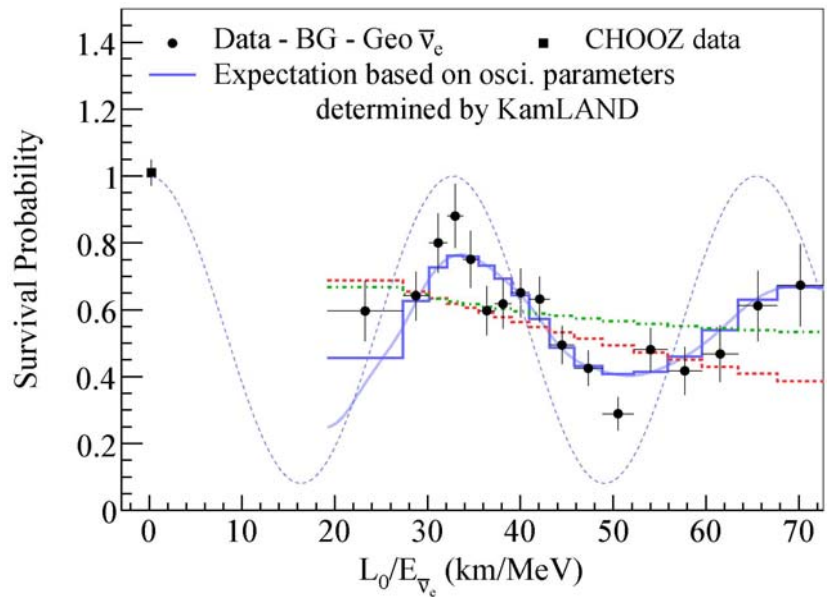

Fig. 27: The same ratio as Fig. 26. The solid (blue), dash (red) and dash-dot (green) histograms are the expectations from the best-fit oscillations, best-fit decay and best-fit decoherence, taking into account the individual time-dependent flux variations of all reactors and detector effects.

The reactor and solar neutrino oscillation data provide a fascinating test on CPT invariance in the neutrino sector. The sensitivity of CPT measurements depends on the frequency of the neutrino

oscillations and difference of the squared neutrino masses [57]

$$\delta CPT \sim \delta m^2/E \leq 10^{-21} \text{ GeV} \qquad (7)$$

This gives a more stringent upper limit on the CPT violation than that in the baryon sector which is the well-known upper limit on the mass difference between $K_0$ and $\overline{K}_0$, $\delta|m(K_0) - m(\overline{K}_0)| < 4.4 \times 10^{-19}$ GeV (90 % C.L.) [58].

## 7. Natural Nuclear Reactor in the Earth's Core : Geo-reactor

It is generally agreed that the Earth core must produce a significant amount of energy, which is necessary to maintain convection in the outer core as well as the magnetic field of the Earth. It is claimed that a significant part of the heat production in the core is due to the presence of Blobs of concentrated uranium (U) that act as fast breeder reactors [59]. We call this type reactor a geo-reactor. Although the geo-reactor is attractive so as to explain the mechanism for flips of the geo-magnetic field, it is not a mainstream theory. However, nobody can rule out the geo-reactor hypothesis by any evidence. Usually the content of the inner core is based on the carbonaceous and oxygen-rich chondrites. As a result, U and thorium (Th) do not sink, but stay in the crust and mantle. On the other hand, in the natural nuclear reactor model, the inner core comprises the rare enstatite chondrites and poor in oxygen. In consequence, U and Th can sink to the Earth's center. Collected evidence shows that there are still open issues with models of the Earth.

KamLAND can perform a fully independent check of the geo-reactor hypothesis and set a clean limit on its possible power output. The signature from the geo-reactor is given as a time-constant $\bar{\nu}_e$ flux over the duration of the experiment. Time evolution of the expected and observed reactor rates at KamLAND for $\bar{\nu}_e$'s with energies between 2.6 and 8.5 MeV shown in Fig. 18(b) is used for this analysis. The observed event rate is plotted at the exposure-weighted expected event rate of reactor $\bar{\nu}_e$'s + backgrounds in Fig. 28(a). The oscillation parameters used to calculate the expected reactor $\bar{\nu}_e$ rate are the best-fit values from the global oscillation analysis, including constraints on $\theta_{13}$ from accelerators and short-baseline reactor experiments (see Table 3). The contribution of geo $\bar{\nu}_e$'s is negligible in this energy range. The intercept can be interpreted as the reactor-independent constant background rate. The geo-reactor signal is embedded in the y-axis intercept. The fit tells that the intercept is consistent with known backgrounds within $1\sigma$ and gives a limit on the geo-reactor power of < 3.1 TW at 90 % C.L. and < 3.7 TW at 95 % C.L. This is comparable to 3 TW of theoretical predictions for geo-reactor. Therefore with a new set of data and more extensive study, KamLAND might be able to set a limit on theoretical predictions.

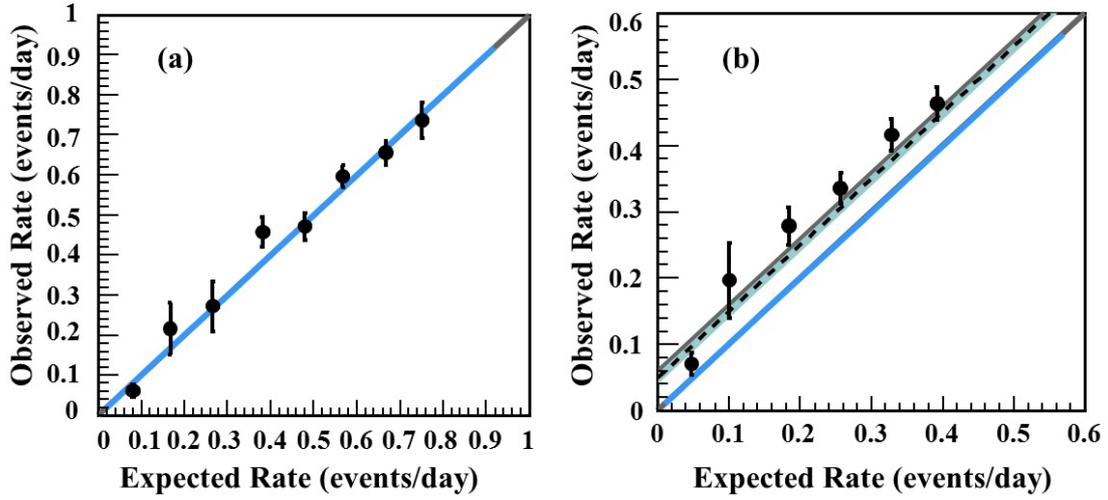

Fig. 28. Observed $\bar{\nu}_e$ event rate versus expected reactor $\bar{\nu}_e$ + background rate with energies of (a) 2.6 MeV < $E_{prompt}$ < 8.5 MeV and (b) 0.9 MeV < $E_{prompt}$ < 2.6 MeV.

The same plot as Fig. 28(a) for the prompt energies between 0.9 MeV and 2.6 MeV is shown in Fig. 28(b), using the data sample of Fig. 19(a). All data points exceed over the expected rates from reactor $\bar{\nu}_e$ and background events (blue color), but are consistent with the expected rates from reactor $\bar{\nu}_e$ + background + geo $\bar{\nu}_e$ events (dashed line). This result demonstrates the geoneutrino detection at KamLAND.

## 8. Geoneutrino Detection

Thanks to a 1000 ton large target volume, KamLAND has a first chance to search for the geo-neutrinos ($\bar{\nu}_e$'s) produced from the $^{238}$U and $^{232}$Th decay chains [29]. The use of geologically produced antineutrinos to study Earth science was first introduced by Marx [60][61], Markov[62] and Eder [63] in 1960's, and then reviewed several times by several authors[64-68]

Radiogenic heat dominantly from decays of $^{238}$U, $^{232}$Th and $^{40}$K is supposed to contribute approximately half of the total measured heat dissipation rate from the Earth which is 44.2 ± 1.0 TW ($10^{12}$ Watt) or 31 ± 1 TW, depending on an analysis procedure [69][70]. Another 1/2 is believed to come from Primordial energy on the planetary accretion and latent heat of core solidification. Heat generation of the Earth is the basic factor to understand the interior dynamics of plate tectonics, mantle convection and terrestrial magnetism. More fundamentally why such heat exists at the present Earth asks how our Earth was born and has been evolving. Detecting geo $\bar{\nu}_e$'s from radioactive elements in the Earth is expected to bring direct insight from the deep Earth and is essential to study the above fundamental issues.

The $^{238}$U and $^{232}$Th decays via a series of well-established $\alpha$– and $\beta$- processes emit 6 and 4 $\bar{\nu}_e$'s, respectively. The $^{40}$K decays through two branching modes, 89.28 % of a $\beta$- decay and 10.72 % of an electron capture, accompanying $\bar{\nu}_e$ or $\nu_e$. The expected $\bar{\nu}_e$ energy distributions of these decay chains are shown in Fig. 29. KamLAND detects $\bar{\nu}_e$'s with $E > 1.8$ MeV due to the reaction threshold energy of the inverse $\beta$-decay, resulting to be sensitive to $\bar{\nu}_e$'s from $^{238}$U and $^{232}$Th.

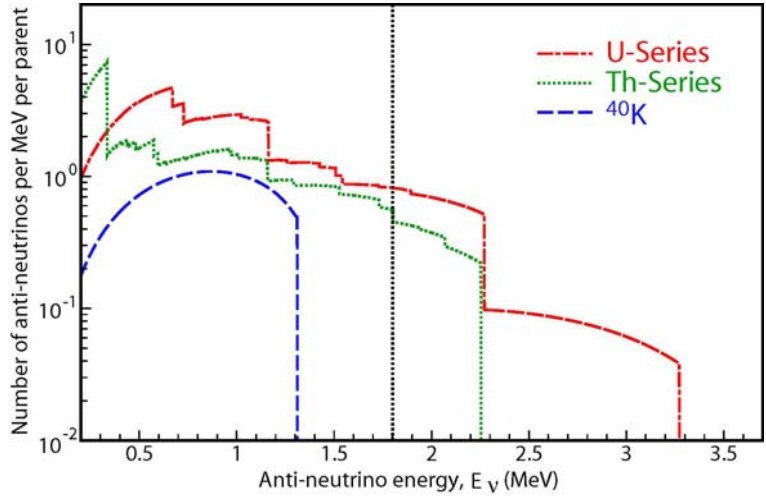

Fig. 29. Energy distributions of the expected $\bar{\nu}_e$'s from $^{238}$U, $^{232}$Th and $^{40}$K decay chains. The vertical black dotted line indicates $E = 1.8$ MeV, the threshold energy of inverse $\beta$-decay

To estimate geo $\bar{\nu}_e$ flux at the Earth surface, we constructed a reference Earth model including the currently available geophysical and geochemical knowledge [38]. Seismic data provides the structural feature of the inner Earth which divides the Earth into sediment, crust, mantle and core. These regions are further sub-divided as listed in Table 4. The seismological analysis also yields the thickness maps of sediment and crust at $2° \times 2°$ resolution and the global thicknesses of core and mantle. Each sub-region has different $^{238}$U and $^{232}$Th concentrations. The bulk chemical composition of the Earth is studied based on the analysis of chondritic meteorite which is thought to be close to the Earth ingredients, and then the bulk silicate Earth model (BSE model [71]) was constructed. Our reference Earth model completely follows the BSE model. Table 4 summarizes the $^{238}$U and $^{232}$Th concentrations of our reference Earth model. One can see the $^{232}$Th/$^{238}$U mass ratio distributes between 2.2 and 4.0 in each region. The present model assumes that the ratio of chemical composition, $^{232}$Th/$^{238}$U at each region is uniform, and $^{238}$U and $^{232}$Th are absent inside the core.

|  |  | U (ppm) | U series $\bar{\nu}_e$ | Th (ppm) | Th series $\bar{\nu}_e$ |
|---|---|---|---|---|---|
| Sediment | continental | 2.8 | $6.11 \times 10^4$ | 10.7 | $5.07 \times 10^4$ |
|  | oceanic | 1.68 | $1.35 \times 10^4$ | 6.91 | $1.20 \times 10^4$ |
| Continental Crust | upper | 2.8 | $1.15 \times 10^6$ | 10.7 | $9.57 \times 10^5$ |
|  | middle | 1.6 | $4.31 \times 10^5$ | 6.1 | $3.57 \times 10^5$ |
|  | lower | 0.2 | $5.25 \times 10^4$ | 1.2 | $6.85 \times 10^4$ |
| Oceanic Crust |  | 0.10 | $9.04 \times 10^3$ | 0.22 | $4.33 \times 10^3$ |
| Mantle | upper | 0.012 | $2.20 \times 10^5$ | 0.048 | $1.91 \times 10^5$ |
|  | lower | 0.012 | $4.03 \times 10^5$ | 0.048 | $3.51 \times 10^5$ |
| Core | outer | 0 | 0 | 0 | 0 |
|  | inner | 0 | 0 | 0 | 0 |

Table 4. $^{238}$U and $^{232}$Th concentrations from the reference Earth model [38], and geo $\bar{\nu}_e$ fluxes at Kamioka in units of cm$^{-2}$sec$^{-1}$.

The geo $\bar{\nu}_e$ flux at a position $\vec{r}$ is determined from the isotopic abundance,

$$\frac{d\Phi(E_\nu,\vec{r})}{dE_\nu} = \sum_i^{U,Th} A_i \frac{dn_i(E_\nu)}{dE_\nu} \int_\oplus d^3\vec{r'} \frac{a_i(\vec{r'})\rho(\vec{r'})P(E_\nu,|\vec{r}-\vec{r'}|)}{4\pi|\vec{r}-\vec{r'}|^2}$$

(8),

where $A_i$ is the decay rate per unit mass, the integration extends over the Earth's volume, $a_i(\vec{r'})$ is the isotope mass per unit rock mass, $dn_i(E_\nu)/dE_\nu$ is the energy spectrum of neutrinos for each mode of decays, $\rho(\vec{r'})$ is the rock density, and $P(E_\nu,|\vec{r}-\vec{r'}|)$ is the $\bar{\nu}_e$ survival probability after travelling a distance $|\vec{r}-\vec{r'}|$. This probability derives from the now accepted phenomenon of neutrino oscillation. The expected geoneutrino flux from each region at KamLAND (36.42° N, 137.31° E), including a suppression factor due to neutrino oscillations is also shown in Table 4. A total geo $\bar{\nu}_e$ flux is 2.34 $\times 10^6$ cm$^{-2}$ sec$^{-1}$ from the $^{238}$U and 1.98 $\times 10^6$ cm$^{-2}$ sec$^{-1}$ from $^{232}$Th decay chains in which the sediment, crust and mantle contributions are 3 %, 70 % and 27 %, respectively. The effect of local geology and specific structure of Japan Island Arc was found to be less than 10 % error on the total expected flux. Approximately 25 % and 50 % of the total flux originates within 50 km and 500 km of KamLAND, respectively, which shows that a large fraction of the expected geo $\bar{\nu}_e$ flux originates in the area surrounding KamLAND.

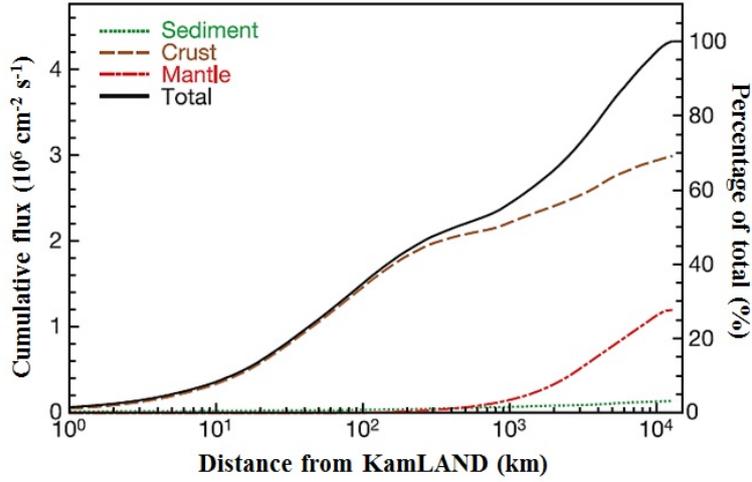

Fig. 30. The expected total $^{238}$U and $^{232}$Th geo $\bar{\nu}_e$ flux within a given distance from KamLAND.

**8.1 Data Analysis**

The data sample based on a total detector live-time of 749.1 ± 0.5 days taken in March 7, 2002 to October 30, 2004, was used to search for geo $\bar{\nu}_e$'s [72]. The $\bar{\nu}_e$ event cuts are almost the same as the reactor $\bar{\nu}_e$ cuts for the ANA-II data sample. However to reduce accidental coincidence events, the fiducial volume was squeezed to 5 m spherical radius. The number of target protons is estimated at (3.46 ± 0.17)×10$^{31}$ on the basis of target proton density and a spherical fiducial scintillator volume, resulting in a total exposure of (7.09 ± 0.35)×10$^{31}$ target proton years. Also, the stringent time and space correlations between prompt and delayed events, 0.5 μsec < $\Delta T$ < 500 μsec and $\Delta R$ < 100 cm were applied to qualify the inverse $\beta$-decay events as shown in Fig. 31. The overall efficiency for detecting geo $\bar{\nu}_e$ candidates with energies between 1.7 MeV and 3.4 MeV in the fiducial volume is estimated to be 0.687 ± 0.007. The energy range reaches below the inverse β-decay threshold of 1.8 MeV owing to the detector energy resolution.

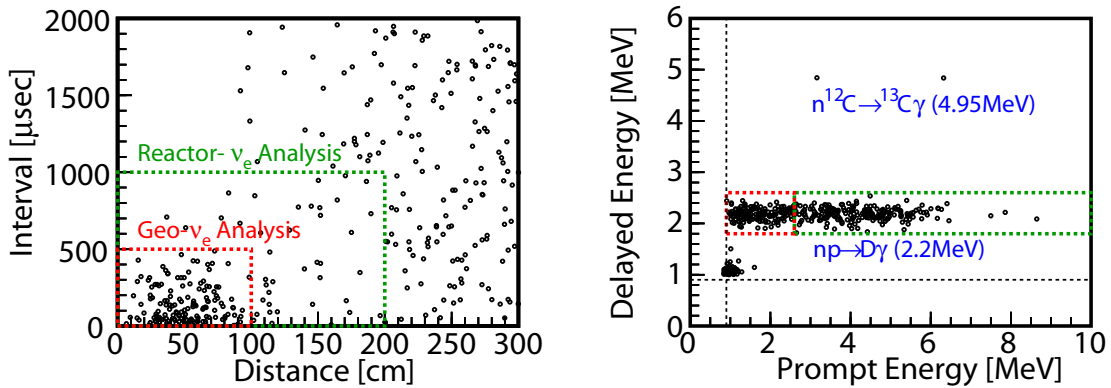

Fig. 31. (Left panel) $\Delta R$ - $\Delta T$ plot, and (Right panel) the prompt and delayed energy correlation of $\bar{\nu}_e$ candidate events. Points inside the dotted red boxes are used for the geo $\bar{\nu}_e$ analysis.

The total number of observed $\bar{\nu}_e$ candidates is 152, with their energy distribution shown in Fig. 32. On the other hand, backgrounds are dominated by reactor $\bar{\nu}_e$'s, and by α-particle induced neutron backgrounds from the $^{13}C(\alpha, n)$ $^{16}O$ reaction. The energy spectrum of reactor $\bar{\nu}_e$'s in this energy region is determined by analyzing $\bar{\nu}_e$'s with energies greater than 3.4 MeV, where there is no signal from the geo $\bar{\nu}_e$'s. Using the reactor neutrino oscillation parameters, the number of reactor $\bar{\nu}_e$ background events is estimated to be 80.4 ± 7.2. The number of α-particle induced neutron background events is 42 ± 11. Including other small contribution from random coincidences, the total background is 127 ± 13 events (1σ error). Thus $25^{+19}_{-18}$ geo $\bar{\nu}_e$ candidates from the $^{238}U$ and $^{232}Th$ decay chains are extracted. This result is consistent with the 19 events predicted by our reference Earth model. Dividing by the detection efficiency, live-time, and number of target protons, the total geo $\bar{\nu}_e$'s detected rate is obtained $5.1^{+3.9}_{-3.6} \times 10^{-31}$ $\bar{\nu}_e$ per target proton per year.

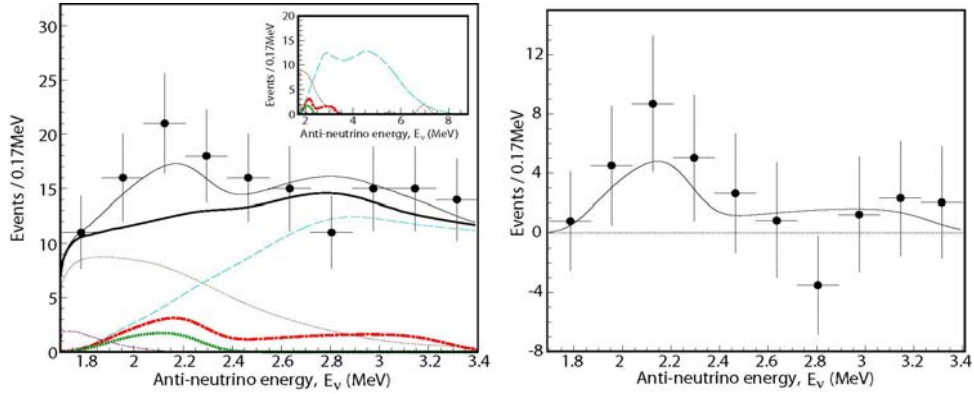

Fig. 32. (Left panel) $\bar{\nu}_e$ energy spectra of the candidate events (data), the total expectation (thin solid black line), the total backgrounds (thick solid black line), the expected $^{238}U$ signals (dot-dashed red line), the expected $^{232}Th$ signals (dotted green line), and the backgrounds due to reactor $\bar{\nu}_e$ (dashed blue line), $^{13}C$ $(\alpha, n)$ $^{16}O$ reactions (dotted brown line) and random coincidences (dot-dashed blue line). (Right panel) $\bar{\nu}_e$ energy spectra of the candidate events subtracted by the total backgrounds.

An unbinned maximum likelihood analysis of the $\bar{\nu}_e$ energy spectrum between 1.7 MeV and 3.4 MeV is carried out as a cross-check of extracting the number of geo $\bar{\nu}_e$ events. Here the reactor neutrino oscillation parameters are allowed to adopt the values of the best-fit ±1σ. The confidence intervals for the number of geo $\bar{\nu}_e$'s are shown in Fig. 33 (Left). The best fit gives 21 geo $\bar{\nu}_e$'s shown by the dark circle. Although this result is somehow contradict with that of the reference Earth model indicated by the rectangular box in Fig. 33 (Left), even 68.3 % C.L. contour covers this box. Based on a study of chondritic meteorites, the Th/U mass ratio in the Earth is believed to be between 3.7 and 4.1, and is known better than either absolute concentration. Hence we investigate the $\chi^2$ behavior, assuming a Th/U mass ratio of 3.9, which corresponds to the $\chi^2$ distribution along the dotted line in Fig. 33 (Right). With the 90 % C.L. the total number of $^{238}U$ and $^{232}Th$ geo $\bar{\nu}_e$'s are 4.5 to 54.2. The central value of 28.0 is consistent with 25 obtained in the above rate-only analysis. The 99 % confidence upper limit of the total $^{238}U$ and $^{232}Th$ geo $\bar{\nu}_e$ flux at KamLAND is $1.62 \times 10^7$ cm$^{-2}$ sec$^{-1}$ which corresponds to an upper limit on radiogenic power of 60 TW for the reference Earth model.

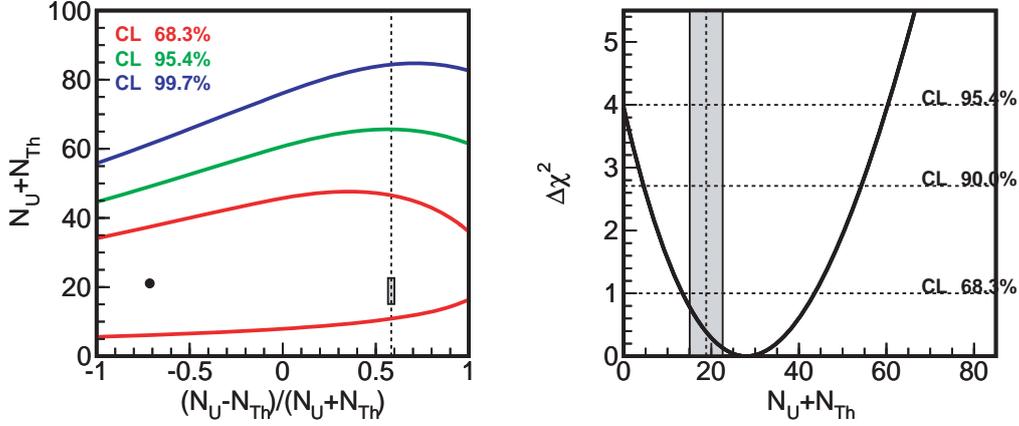

Fig. 33. Confidence intervals of geo $\bar{\nu}_e$'s. (Left panel) Dark circle and rectangular area stand for the best-fit and the prediction of the reference Earth model. (Right panel) The vertical dotted line and gray band give the constraint from the reference Earth model.

The Borexino collaboration at Gran Sasso reported an excess attributed to geo $\bar{\nu}_e$'s [73]. While Borexino confirmed the excess from KamLAND, its result was not precise enough to significantly constrain geophysical models. The KamLAND second results on a measurement of geo $\bar{\nu}_e$ flux were presented based on data collected from March 9, 2002 to November 4, 2009, which includes the data used in ANA-I, ANA-II and ANA-III [74]. The total exposure to $\bar{\nu}_e$ is $3.49 \times 10^{32}$ target-proton years, a fivefold improvement over the first KamLAND geo $\bar{\nu}_e$ result [72]. The expected signal based on the reference Earth model increased from 19 events to 106 events. After the first results, several improvements to reduce systematic errors for selecting $\bar{\nu}_e$ events were carried out as mentioned above in the ANA-III data analysis. The overall efficiencies for U and Th events are estimated to be 80.7 % and 75.1 %, respectively. The difference comes from the energy-dependent efficiencies.

We observed 841 candidate events between 0.9 MeV and 2.6 MeV; the expected number is 729.4 ±32.3 including reactor $\bar{\nu}_e$'s. Interpreting the excess as the geo $\bar{\nu}_e$ implies $111^{+45}_{-43}$ events. The statistical significance is 99.55 %. The significance of excess is better determined from an unbinned maximum likelihood fit taking account of event rate, energy, and time information in the energy range $0.9 < E_{\text{prompt}} < 8.5$ MeV, i.e. the simultaneous analysis of the geo and reactor $\bar{\nu}_e$'s including the effect of neutrino oscillations. Fig. 34 shows the prompt energy spectrum of candidate events and the fitted backgrounds. The resulting best-fit values are 65 and 33 geo $\bar{\nu}_e$ events from $^{238}$U and $^{232}$Th, respectively, with confidence intervals shown in Fig. 35(a). The result is consistent with the reference Earth model. Assuming a mass ratio of Th to U of 3.9 based on abundances found in chondritic meteorites, the total number of excess events is $106^{+29}_{-28}$ from geo $\bar{\nu}_e$'s, as shown in Fig. 35(b), corresponding to a flux of $4.3^{+1.2}_{-1.1} \times 10^6$ cm$^{-2}$s$^{-1}$. From the $\Delta\chi^2$-profile, the null hypothesis, the absence of $\bar{\nu}_e$ events ($N_U + N_{Th} = 0$) is rejected at 99.997 % C.L.

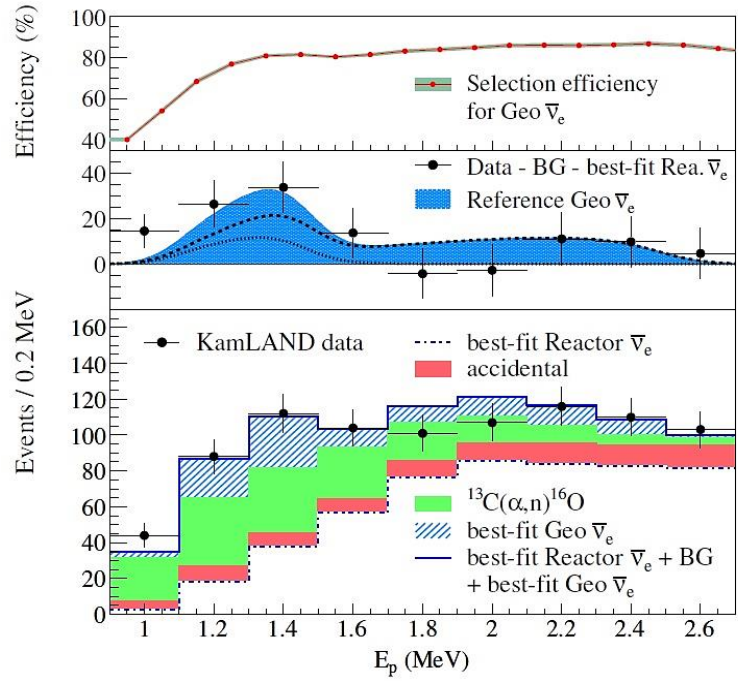

Fig. 34. Prompt energy spectrum of the $\bar{\nu}_e$ events in the low-energy region. (Bottom panel) Data together with the fitted background and geo $\bar{\nu}_e$ contributions. The shaded background and geo $\bar{\nu}_e$ histograms are cumulative. (Middle panel) Background and reactor $\bar{\nu}_e$ subtracted data together with the geo $\bar{\nu}_e$'s for the decay chains of U (dashed) and Th (dotted) calculated from the reference Earth model [38]. (Top panel) The energy-dependent selection efficiency.

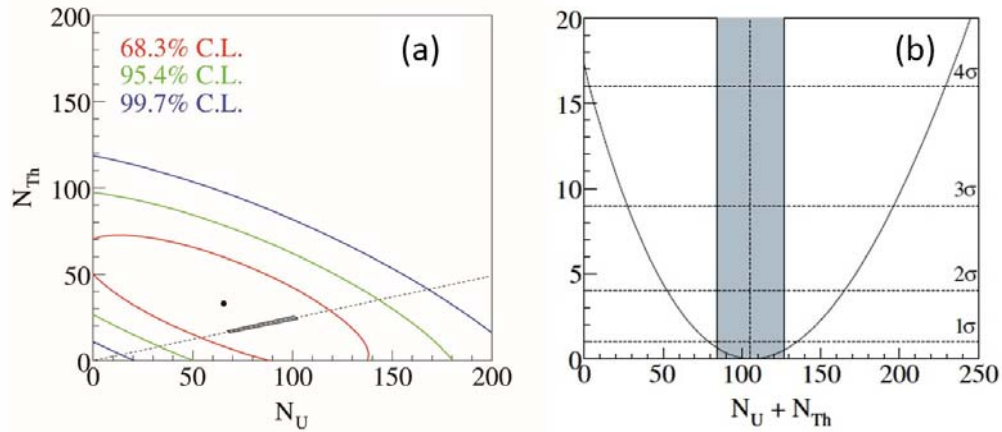

Fig. 35. (a) 68.3 %, 95.4 % and 99.7 % C.L. contours for the number of detected U and Th geo $\bar{\nu}_e$ events. The dot represents the best-fit value. The small shaded region represents the expectation from the reference Earth model. The dashed line represents the expectations from the mass ratio, Th/U = 3.9, obtained from chondritic meteorite abundances. (b) $\Delta\chi^2$ – profile of the total number of detected U and Th geo $\bar{\nu}_e$ events, fixing the mass ratio to the chondritic meteorite constraint. The BSE geological model [71] predicts the value of $N_U + N_{Th}$ shown in the gray band.

For the geo $\bar{\nu}_e$ flux in the ANA-IV data, we incorporate all available constraints on oscillation parameters. Fig. 36 shows the measured geo $\bar{\nu}_e$ event spectrum after subtracting the best-fit reactor $\bar{\nu}_e$ and background spectra. Assuming a Th/U mass ratio of 3.9, the total number of U and Th geo $\bar{\nu}_e$ events is $116^{+28}_{-27}$, which corresponds to the $\bar{\nu}_e$ flux of $3.4^{+0.8}_{-0.8} \times 10^6$ cm$^{-2}$s$^{-1}$ at KamLAND. The $\Delta\chi^2$ profile shows that the null hypothesis is rejected with 99.998 C.L.

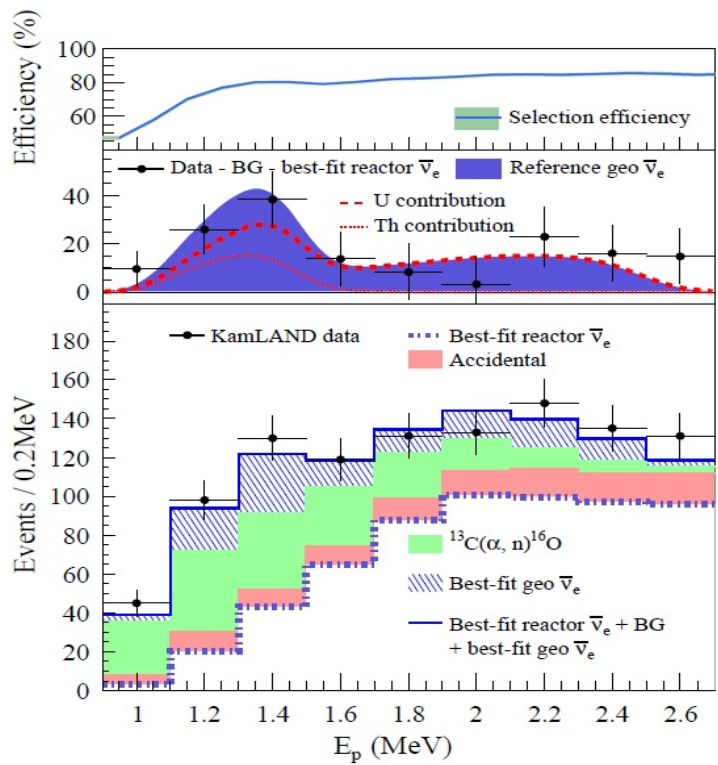

Fig. 36. The caption is the same as that of Fig. 34.

## 8.2 Constraints on Earth Models.

While the mantle is the most massive layer of the Earth's interior, its chemical composition is still uncertain. A quantitative estimate of the heat production by radiogenic components is of particular importance for understanding dynamic processes such as mantle convection. Indeed, precisely how the mantle convects is still not fully understood, and controversy remains as to whether two-layer convection or whole-volume convection provides a more accurate description. In this work, we carry out a comparison of existing Earth models using the KamLAND geo $\bar{\nu}_e$ data on the basis of simple but appropriate assumptions.

The crustal contribution to the flux at KamLAND can be estimated from compositional data through rock sampling [38]. Since current Earth models predict that the lithophiles U and Th are absent in the core, to a first approximation of the radiogenic heat, we attribute any excess above the crustal contribution to U and Th uniformly distributed throughout the mantle. Under these generic assumptions, the measured KamLAND geo $\bar{\nu}_e$ flux translates to a total radiogenic heat production of $11.2^{+7.9}_{-5.1}$ TW from U and Th. This calculation accounts for crustal uncertainties of 17 % and 10 % for U and Th, respectively, including correlated errors as suggested in [75]. To parameterize the planetary-scale energy balance, the fraction of the global heat production from radioactive decays, the so-called "Urey ratio" is introduced. Allowing for mantle heat contributions of 3.0 TW from other isotope decays [76][77], we find that the convective Urey ratio, the contribution to the Urey ratio from just the mantle, is between 0.09 and 0.42 at 68 % C.L. This range favors models that allow for a substantial but not dominant contribution from the Earth's primordial heat supply.

Several established estimates of the BSE (Bulk Silicate Earth model) composition give different geo $\bar{\nu}_e$ flux predictions. Reference [78] categorizes the models into three groups: geochemical, cosmochemical, and geodynamical. Geochemical models [71], such as the reference Earth model [38], use primordial compositions equal to those found in CI carbonaceous chondrites, but allow for elemental enrichment by differentiation, as deduced from terrestrial samples. Cosmochemical models [79] assume a mantle composition similar to that of enstatite chondrites, and yield a lower radiogenic abundance. Geodynamical models [80], on the other hand, require higher radiogenic abundances in order to drive realistic mantle convection.

The observed geo $\bar{\nu}_e$ flux at KamLAND is compared with the expectations from these BSE composition models assuming a common estimated crustal contribution [38] in Fig. 37. The $\bar{\nu}_e$ flux predictions vary within the plotted vertical bands due to uncertainties in both the abundances of radioactive elements in the mantle as well as their distributions. The spread of the slope reflects the difference between two extreme radiochemical distributions: the "homogeneous hypothesis" in which U and Th are assumed to be distributed uniformly throughout the mantle, and the "sunken-layer hypothesis", which assumes that all of the U and Th below the crust collects at the mantle-core interface. While the statistical treatment of geological uncertainties is not straightforward, assuming Gaussian errors for the crustal contribution and for the BSE abundances, we find that the geodynamical prediction with the homogeneous hypothesis is disfavored at 89 % C.L. However, due to the limited statistical power of the data, all BSE composition models are still consistent within ~2 σ C.L. ]

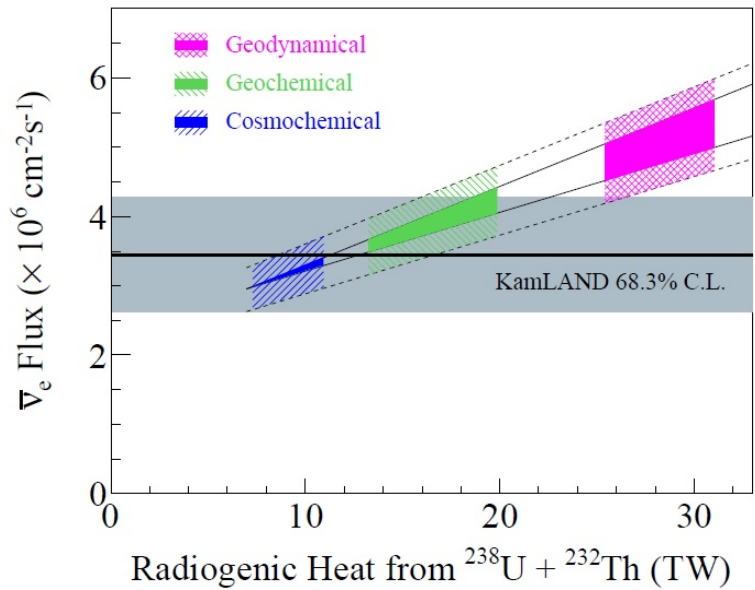

Fig. 37. Geo $\bar{\nu}_e$ flux versus radiogenic heat from the decay chains of $^{238}$U and $^{232}$Th. The measured geo $\bar{\nu}_e$ flux (gray band) is compared with the expectations for the different mantle models from cosmochemical [79], geochemical [71], and geodynamical [80] estimates (color bands). The slope band starting at 7 TW indicates the response to the mantle $\bar{\nu}_e$ flux, which varies between the homogeneous and sunken-layer hypotheses (solid lines), discussed in the text. The upper and lower dashed lines incorporate the uncertainty in the crustal contribution.

**8.3 Further Improvement**

The directional measurement of incoming geo $\bar{\nu}_e$'s provides much information on the heat generation inside the Earth. Fig. 38 shows the zenith angle distribution of the $^{238}$U geo $\bar{\nu}_e$'s at Kamioka calculated by the reference Earth model. It can be seen that the zenith angle distribution gives the crust and mantle components of geo $\bar{\nu}_e$'s. It is also interested to verify a null contribution of the $^{238}$U and $^{232}$Th geo $\bar{\nu}_e$'s in the Earth core which is the basic assumption of the BSE model.

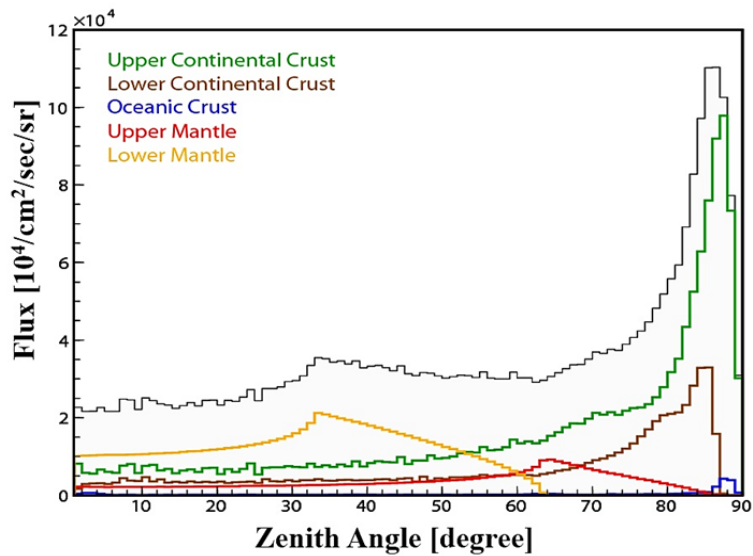

Fig. 38. Zenith angle distributions of the $^{238}$U geo $\bar{\nu}_e$'s at Kamioka produced in the upper/lower continental crust, oceanic crust, and upper/lower mantle. The sum of all contributions is shown by a black histogram.

In detecting $\bar{\nu}_e$'s via the inverse $\beta$-decay process, $\bar{\nu}_e + p \rightarrow e^+ + n$, the incident $\bar{\nu}_e$ direction is approximated to the direction determined by two vertexes of the prompt $e^+$ and the delayed $\gamma$-ray produced by the thermal neutron captured on a proton or on a material loaded inside liquid scintillator. If the $\gamma$-ray production position is well identified, the measured direction shows an enough correlation to the $\bar{\nu}_e$ direction as shown in Fig. 39(Left). However the precise measurement of the delayed $\gamma$-ray production position is too difficult due to multiple scatterings. The direction obtained by reconstructed vertexes of $e^+$ and $\gamma$-ray shows less correlation to that of $\bar{\nu}_e$ (see Fig. 39(Right)). One possibility to solve this problem is to develop a material loaded liquid scintillator which provides delayed $\alpha$-particles and/or $\beta$-rays instead of $\gamma$-rays after capturing thermal neutrons by the loaded materials. Fig. 39 compares the Monte-Carlo angular distributions of $^{10}$B (1.0 wt %) loaded, $^{6}$Li (0.15 wt %) loaded, and KamLAND LS along the incident $\bar{\nu}_e$ direction [81][82]. Here thermal neutron capture signals are generated through (n, α) reactions, n + $^{10}$B (BR = 94 %) → $^{7}$Li* (→ $^{7}$Li + γ) + α, n + $^{10}$B (BR = 6 %) → $^{7}$Li + α, and n + $^{6}$Li → $^{3}$H + α. The $\bar{\nu}_e$ direction measurement is the most urgent task in future geo $\bar{\nu}_e$ experiments.

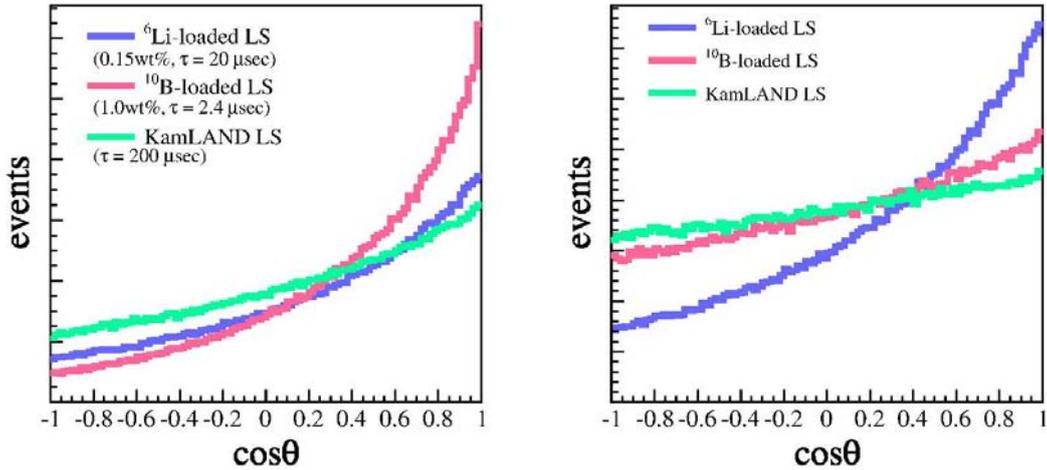

Fig. 39. Angular correlation between the reconstructed and incident $\bar{\nu}_e$ direction for $^{10}$B (1.0 wt %) loaded, $^{6}$Li (0.15 wt %) loaded, and KamLAND liquid scintillators, using Monte Carlo generated events. The reconstructed direction is obtained by the reconstructed vertex of $e^+$'s and the simulated position of γ / α (Left panel), and by the reconstructed vertexes of $e^+$'s and γ / α (Right panel).

## 9. Summary

KamLAND, the first homogeneous large volume LS detector, demonstrated the reactor $\bar{\nu}_e$ disappearance at long baselines for the first time. The observed event-rate suppression and the spectrum-shape distortion conclusively shows that the reactor $\bar{\nu}_e$ disappearance is due to neutrino oscillations. With the assumption of CPT invariance, the Large Mixing Angle (LMA) is the only remaining oscillation solution consistent with the KamLAND result. Furthermore, because observed $\bar{\nu}_e$'s are created in nuclear reactors rather than in the core of the Sun, several alternative possible solutions to the solar neutrino deficit, for instance, the neutrino magnetic moment and unknown neutrino interactions inside the Sun, are excluded. This means that the solar neutrino deficit problem which lasted for almost 30 years was finally solved. The KamLAND data clearly illustrates the oscillatory shape of reactor $\bar{\nu}_e$'s arising from neutrino oscillations, thanks to the concentration of nuclear power stations around the Kamioka mine with 180 km of the flux-weighted average baseline.

The first experimental study of $\bar{\nu}_e$'s from the Earth's interior was performed, using KamLAND. The present observation is in agreement with the predictions from existing the BSE (Bulk Silicate

Earth model) composition models within ~2 σ C.L. The KamLAND result of the geo $\bar{\nu}_e$'s flux provides a deep probe for studying portions of the planet that are otherwise inaccessible to us. In the future, multisite flux data at a combination of crustal and oceanic geological sites enables us to estimate of the crustal contribution from a statistical correlation analysis and constrain mantle abundances more stringently.

## Acknowledgments

The author gratefully acknowledges the colleagues in KamLAND and the generous cooperation of the Kamioka Mining and Smelting Co., and the electric associations in Japan. The KamLAND experiment has been supported by the Japan Society for the Promotion of Science (JSPS), the Japanese Ministry of Education, Culture, Sports, Science and Technology (MEXT), and the United States Department of Energy (DOE).

## References

[1] W. Pauli, A letter to a meeting of physicists in Tubingen (1930).
[2] E. Fermi, Zeits. für Phys. **88**, 161 (1934).
[3] C. Cowan, F. Reines, F. Harrison, H. Kruse and A. McGuire, Science **124**, 103 (1956).
[4] T.D. Lee and C.N. Yang, Phys. Rev. **104**, 254 (1956).
[5] C. Wu et al., Phys. Rev. **105**, 1413 (1957).
[6] B. Pontecorvo, J. Exp. Theor. Phys. **33**, 549 (1957).
[7] M. Goldhaber, L. Grodzins and A. Sunyar, Phys. Rev. **109**, 1015 (1958).
[8] R. Feynman and M. Gell-Mann, Phys. Rev. **109**, 193 (1958).
[9] B. Pontecorvo, JETP **37**, 1751 (1959).
[10] M. Schwartz, Phys. Rev. Lett. **4**, 306 (1960).
[11] G. Danby et al., Phys. Rev. Lett. **9**, 36 (1962).
[12] Z. Maki, M. Nakagawa and S. Sakata, Prog. Theor. Phys. **28**, 870 (1962).
[13] F. Hasert et al., Phys. Lett. **46B**, 138 (1973).
[14] G. Arnison et al., Phys. Lett. **B122**, 103 (1983).
    M. Banner et al., Phys. Lett. **B122**, 476 (1983).
    G. Arnison et al., Phys. Lett. **B126**, 398 (1983).
    P. Bagnaia et al., Phys. Lett. **B129**, 130 (1983).
[15] ALEPH, DELPHI, L3, OPAL, and SLD Collaborations, and LEP Electroweak Working Group, SLD Electroweak Group, and SLD Heavy Flavor Group, Phys. Reports **427**, 257 (2006).
[16] Y. Fukuda et al., Phys. Rev. Lett. **81**, 1562 (1998).
[17] K. Kodama et al., Phys. Lett. **B504**, 218 (2001).
[18] K. Hirata et al., Phys. Rev. Lett. **58**, 1490 (1987).
[19] R. Bionta et al., Phys. Rev. Lett. **58**, 1494 (1987).
[20] K. Hirata et al., Phys. Rev. Lett. **63**, 16 (1989).
[21] R. Davis et al., Phys. Rev. Lett. **20**, 1205 (1968).
[22] K. Hirata et al., Phys. Lett. **B280**, 146 (1992).
[23] Y. Fukuda et al., Phys. Lett. **B335**, 237 (1994).
[24] A. Suzuki et al., Research for Neutrino Science in the Ultra Low-Background Environment, Proposal for Grants-in-Aid for Scientific Research (1996) (in Japanese).
    A. Suzuki, Nucl. Phys. **B**(Proc. Suppl)**77**, 171 (1999).
[25] B. Berger et al., JIST 4, 04017 (2009).

[26]  K. Nakajima et al., Nucl. Instrum. Methods Phys. Res. Sect. **A 569**, 837 (2006).
[27]  K. Schreckenbach et al., Phys. Lett. **B160**, 325 (1985).
     A.A. Hahn et al., Phys. Lett. **B 218**, 365 (1985).
     P. Vogel et al., Phys. Rev. **C24**, 1543 (1981).
[28]  Particle Data Group, Phys. Rev. **D66**, 010001 (2002).
[29]  R. Raghavan et al., Phys. Rev. Lett. **80**, 635 (1998).
[30]  K. Eguchi et al., Phys. Rev. Lett. **90**, 021802 (2003).
[31]  T. Araki et al., Phys. Rev. Lett. **94**, 081801 (2005).
[32]  S. Abe et al., Phys. Rev. Lett. **100**, 221803 (2008).
[33]  A. Gando et al., Phys. Rev. **D88**, 033001 (2013).
[34]  K. K. Sekharan et al., Phys. Rev. **156**, 1187 (1967).
[35]  JENDL, the Japanese Evaluated Nuclear Data Library.
[36]  S. Harissopulos et al., Phys. Rev. **C72**, 062801 (2005).
[37]  A. Gando et al., Phys. Rev. Lett. **110**, 062502 (2013).
[38]  S. Enomoto et al., Earth Planet. Sci. Lett. **258** 147 (2007).
[39]  G.L. Fogli et al., Phys. Rev. **D66**, 053010 (2002).
[40]  M. Apollonio et al., Phys. Lett. **B466**, 415 (1999).
[41]  F. Boehm et al., Phys. Rev. **D64**, 112001 (2001).
[42]  S.N. Ahmed et al., Phys. Rev. Lett. **92**, 181301 (2004).
[43]  B. Aharmim et al., Phys. Rev. **C72**, 055502 (2005).
[44]  J.N. Bahcall, et al., Astrohys. J. **621**, L85 (2005).
[45]  B.T. Cleveland et al., Astrohys. J. **496**, 505 (1998).
[46]  J.N. Abdurashitov et al., Phys. Rev. **C80**, 015807 (2009).
[47]  G. Bellini et al., Phys. Rev. Lett. **107**, 141302 (2011).
[48]  J. Hosaka et al., Phys. Rev. **D73**, 112001 (2006).
[49]  B. Aharmim et al., arXiv:hep-ph/ 0501216v4.
[50]  K. Abe et al., Phys. Rev. Lett. **107**, 041801 (2011).
[51]  P. Adamson et al., Phys. Rev. Lett. **107**, 181802 (2011).
[52]  Y. Abe et al., Phys. Rev. **D86**, 052008 (2012).
[53]  F.P. An et al., Chinese Phys. **C37**, 011001 (2013).
[54]  J.K. Ahn et al., Phys. Rev. Lett. **108**, 191802 (2012).
[55]  V.D. Barger et al., Phys. Rev. Lett. **82**, 2640 (1999).
[56]  E. Lisi et al., Phys. Rev. Lett. **85**, 1166 (2000).
[57]  J.N. Bahcall, et al., Phys. Lett. **B534**, 120 (2002).
[58]  D.E. Groom, et al., Eur. Phys. J. **C15**, 1 (2000).
[59]  D.F. Hollenbach and J.M. Herndon, Proc. Natl. Acad. Sci. USA (2001).
[60]  G. Marx and N. Menyard, Mitteilungen der Sternwarte, Budapest (1960).
[61]  G. Marx, Czech. J. Phys. **B19**, 1471 (1969).
[62]  M.A. Markov, Neutrino 1964, Nauka, Russia.
[63]  G. Eder, Nucl. Phys. **78**, 657 (1966).
[64]  C. Avilez et al., Phys. Rev. **D23**, 1116 (1981).
[65]  L.M. Kruss et al., Nature **310**, 191 (1984).
[66]  M. Kobayashi and Y. Fukao, Geophys. Res. Lett. **18**, 633 (1991).
[67]  C.G. Rothchild et al., Geophys. Res. Lett. **25**, 1083 (1998).
[68]  F. Mantovani et al., Phys. Rev. **D69**, 013001 (2004).
[69]  H.N. Pollack, et al., Rev. Geophys. **31**, 267 (1993).
[70]  A.M. Hofmeistar and R.E. Criss, Tectonophysics **395**, 159 (2005).
[71]  W.F. McDonough and S. Sun, Chem. Geol. **120**, 223 (1995).
[72]  T. Araki et al., Nature **436**, 499 (2005) .
[73]  G. Bellini et al., Phys. Lett. **B687**, 299 (2010).

[74] A. Gando et al., Nature Geoscience **4**, 647 (2011).
[75] G.L. Fogli et al., Earth, Moon, and Planets **99**, 111 (2006).
[76] R. Arevalo et al., Earth and Planet. Sci. Lett. **278**, 361 (2009).
[77] S. Enomoto, Earth, Moon, and Planets **99**, 131 (2006).
[78] O. Sramek et al., Earth and Planet. Sci. Lett. **361**, 356 (2013).
[79] M. Javoy et al., Earth and Planet. Sci. Lett. **293**, 259 (2010).
[80] D.L. Turcotte and G. Shubert, Geodynamics, Applications of Continuum Physics to Geological Problems, second ed. (Cambridge Univ. Press, Cambridge, 2002).
[81] I. Shimizu, Nucl. Phys **B** (Proc. Suppl.) **168**, 147 (2007).
[82] H.K.M. Tanaka and H. Watanabe, Science Reports **4**, 4708 (2014).